\documentclass[%
  aip,%
 amsmath,amssymb,floatfix,
reprint,%
twocolumn,%
jcp,%
longbibliography,
]{revtex4-1}

\usepackage{amsmath,amsthm,latexsym,amssymb,amsfonts,epsfig}

\usepackage[utf8]{inputenc}
\usepackage{graphicx, texdraw}
\usepackage{color}
\usepackage{bm}

\usepackage{type1ec}
\usepackage{mathtools}
\usepackage{mathrsfs}

\usepackage[outer=1.5cm, inner = 1.5cm, top= 2.05cm, bottom = 2.1cm]{geometry}  % 1.45
\usepackage{enumerate}

\usepackage[colorlinks=true,citecolor=OliveGreen, linkcolor=blue]{hyperref}

\usepackage{csquotes}

\usepackage{color}
\usepackage[svgnames,dvipsnames]{xcolor}

\usepackage{amsmath}
\usepackage{amsfonts}
\usepackage{amssymb}

\usepackage{makecell}

\newcommand{\vect}[1]{\boldsymbol{#1}}

\newcommand{\bNabla}{\boldsymbol{\nabla}}

\newcommand{\R}{\vect{r}}

\newcommand{\e}{\vect{\hat{e}}}

\newcommand{\tauS}{\tau_\mathrm{S}}
\newcommand{\tauB}{\tau_\mathrm{B}}

\newcommand{\kS}{\kappa_\mathrm{S}}
\newcommand{\kB}{\kappa_\mathrm{B}}

\newcommand{\Intd}{\mathrm{d}}

\newcommand{\betaB}{\beta_{\mathrm{B}}}

\newcommand{\bG}{\vect{G}}

\newcommand{\review}[1]{#1}

\usepackage{setspace}
\usepackage{multirow}

\usepackage{changes}

\begin{document}

\title{Frequency-dependent higher-order Stokes singularities near a planar elastic boundary: implications for the hydrodynamics of an active microswimmer near an elastic interface}

\author{Abdallah Daddi-Moussa-Ider}
\email{abdallah.daddi.moussa.ider@uni-duesseldorf.de}
\affiliation
{Institut f\"{u}r Theoretische Physik II: Weiche Materie, Heinrich-Heine-Universit\"{a}t D\"{u}sseldorf, Universit\"{a}tsstra\ss e 1, 40225 D\"{u}sseldorf, Germany}

\author{Christina Kurzthaler}
\affiliation
{Department of Mechanical and Aerospace Engineering, Princeton University, Princeton, NJ 08544, USA}

\author{Christian Hoell}
\affiliation
{Institut f\"{u}r Theoretische Physik II: Weiche Materie, Heinrich-Heine-Universit\"{a}t D\"{u}sseldorf, Universit\"{a}tsstra\ss e 1, 40225 D\"{u}sseldorf, Germany}

\author{Andreas Z\"{o}ttl}
\affiliation{Institute for Theoretical Physics, Technische Universit\"{a}t Wien,
Wiedner Hauptstra\ss e 8-10, 1040 Wien, Austria}

\author{Mehdi Mirzakhanloo}
\affiliation
{Department of Mechanical Engineering, University of California, Berkeley, CA 94720, USA}

\author{Mohammad-Reza Alam}
\affiliation
{Department of Mechanical Engineering, University of California, Berkeley, CA 94720, USA}

\author{Andreas M. Menzel}
\affiliation
{Institut f\"{u}r Theoretische Physik II: Weiche Materie, Heinrich-Heine-Universit\"{a}t D\"{u}sseldorf, Universit\"{a}tsstra\ss e 1, 40225 D\"{u}sseldorf, Germany}

\author{Hartmut L\"{o}wen}
% \email{hartmut.loewen@uni-duesseldorf.de}
\affiliation
{Institut f\"{u}r Theoretische Physik II: Weiche Materie, Heinrich-Heine-Universit\"{a}t D\"{u}sseldorf, Universit\"{a}tsstra\ss e 1, 40225 D\"{u}sseldorf, Germany}

\author{Stephan Gekle}
% \email{stephan.gekle@uni-bayreuth.de}
\affiliation
{Biofluid Simulation and Modeling, Theoretische Physik VI, Universit\"at Bayreuth, Universit\"{a}tsstra{\ss}e 30, 95440 Bayreuth, Germany}

\begin{abstract}
 The emerging field of self-driven active particles in fluid environments has recently created significant interest in the biophysics and bioengineering communities owing to their promising future biomedical and technological applications. These microswimmers move autonomously through aqueous media where under realistic situations they encounter a plethora of external stimuli and confining surfaces with peculiar elastic properties. Based on a far-field hydrodynamic model, we present an analytical theory to describe the physical interaction and hydrodynamic couplings between a self-propelled active microswimmer and an elastic interface that features resistance toward shear and bending. We model the active agent as a superposition of higher-order Stokes singularities and elucidate the associated translational and rotational velocities induced by the nearby elastic boundary. Our results show that the velocities can be decomposed in shear and bending related contributions which approach the velocities of active agents close to a no-slip rigid wall in the steady limit. The transient dynamics predict that contributions to the velocities of the microswimmer due to bending resistance are generally more pronounced than to shear resistance. Bending can enhance (suppress) the velocities resulting from higher-order singularities whereas the shear-related contribution decreases (increases) the velocities. Most prominently, we find that near an elastic interface of only energetic resistance toward shear deformation, such as that of an elastic capsule designed for drug delivery, a swimming bacterium undergoes rotation of the same sense as observed near a no-slip wall. In contrast to that, near an interface of only energetic resistance toward bending, such as that of a fluid vesicle or liposome, we find a reversed sense of rotation. Our results provide insight into the control and guidance of artificial and synthetic self-propelling active microswimmers near elastic confinements. 
\end{abstract}
\date{\today}

\maketitle

\section{Introduction}
Artificial nano- and microscale machines hold great potential for future biomedical applications such as precision nanosurgery, biopsy, or transport of radioactive substances to tumor sites~\cite{wanggao12, wang13, paxton04}. These active particles have the ability to move autonomously in biofluids and could reach inaccessible areas of the body to perform delicate and precise tasks. Recent advances in the field have provided a fundamental understanding of various physical phenomena arising in active matter systems~\cite{lauga09,marchetti13, elgeti15, menzel15, bechinger16, zottl16, lauga16ARFM, illien17, dey19}, which exhibit strikingly different behavior than their passive counterparts.
Suspensions of active agents display fascinating collective behavior and unusual spatiotemporal patterns, including propagating density waves~\cite{gregoire04, mishra10, menzel12}, motility-induced phase separation~\cite{tailleur08, palacci13, buttinoni13, speck14, speck15}, and the emergence of active turbulence~\cite{wensink12pnas, wensink12, dunkel13, heidenreich14, kaiser14, heidenreich16}.

While passive particles can be set into motion under the action of an external field, active particles self-propel by converting energy from their environment into mechanical work. At low Reynolds numbers, microswimmers have to employ effective self-propulsion mechanisms that break the time-reversal symmetry of the Stokes flow~\cite{lauga09}, a property commonly referred to as Purcell's scallop theorem~\cite{purcell77,stone96,lauga11prl,lauga11}. For instance, many biological microswimmers perform a non-reciprocal deformation cycle of their body via, e.g., rotating flagella or beating cilia~\cite{lighthill52, blake71cilia, brennen77, sareh13}, whereas synthetic microswimmers move via phoretic effects caused by their asymmetric surface properties~\cite{golestanian05, golestanian07, howse07, liebchen16, sharifi13, michelin14, popescu16, Kurzthaler:2018, varma18}, or by non-reciprocal deformation of their shape~\cite{najafi04, najafi05, dreyfus05, golestanian08, snezhko09, ledesma12, klotsa15, grosjean16, nasouri17,mirzakhanloo18scirep, nasouri19, sukhov19}. 

In many biologically relevant situations, motion occurs in the presence of surfaces that significantly modify the hydrodynamic flows and thereby strongly affect the transport properties, function, and survival of suspended particles and microorganisms.
Confining boundaries play an important role in many engineering and biological processes ranging from the rheology of colloidal suspensions~\cite{lowen94, isa09, klapp08} to the transport of nanoparticles and various molecules through micro- and nanochannels~\cite{stavis05, huh07}. Moreover, microswimmers encounter in their natural habitats a plethora of different types of surfaces with various geometric and elastic properties. Examples include sperm cells in the female reproductive tract~\cite{Nosrati:2015}, bacterial pathogens in microvasculature channels~\cite{Moriarty:2008}, or bacteria in biofilms~\cite{Mazza:2016}. Thus, surface-related effects on their motility may entail important consequences for a large number of biological systems, including biofilm formation, bacterial adhesion, and microbial activity~\cite{van-loosdrecht90, o-toole00}.

Transport properties of active agents near a no-slip rigid planar wall reveal various interesting features~\cite{zargar09, ishimoto13, ishimoto14, li14, uspal15, ibrahim15, schaar15, das15, mozaffari16, simmchen16, lintuvuori16, lushi17, ishimoto17, ruhle18, mozaffari18, daddi18, shen18, mirzakhanloo18}, including their escape from the wall, a stationary hovering state, or gliding along the boundary maintaining a constant orientation during their navigation.
Interestingly, flagellated bacteria display circular swimming trajectories close to surfaces as a consequence of hydrodynamic couplings~\cite{lauga06}. Their swimming direction can be qualitatively influenced by the nature of the boundary conditions at the interface such that, e.g., the circular motion is reversed at a free air-liquid interface when compared to a no-slip wall~\cite{lopez14}.
Bacterial swimming in the close vicinity of a boundary has been addressed theoretically using a two-dimensional singularity model combined with a complex variable approach~\cite{crowdy11}, a resistive force theory~\cite{di-leonardo11}, or a multipole expansion technique~\cite{lopez14}. Further, it has been shown that the presence of a nearby wall can lead to a change in the waveform assumed by actuated flagella causing a strong alteration of the resulting propulsive force~\cite{evans10}.
Under applied shear flow, swimming bacteria~\cite{hill07, kaya12, altshuler13, rusconi14, figueroa15,mathijssen16prl, mathijssen19} and sperm cells~\cite{bretherton61, kantsler14, tung15} near surfaces may inhibit their circular motion and exhibit rheotaxis leading to motion against imposed shear flow. Likewise, the rheotactic behavior of a self-diffusiophoretic particle has been investigated numerically by means of boundary integral simulations~\cite{uspal15rheo}. Direct measurements of the flow field generated by individual swimming \emph{E. coli} both far from and near a solid surface have revealed the relative importance of fluid dynamics and rotational diffusion in bacterial locomotion~\cite{drescher11}. 
More recently, it has been shown that \emph{E. coli} bacteria use transient adhesion to nearby surfaces as a generic mechanism to regulate their motility and transport properties in confinements~\cite{ipina19}.
Remarkably, a nearby wall alone can enable self-phoresis of homogeneous and isotropic active particles~\cite{yariv16}. The behavior of self-propelled nano- and micro-rods in a channel has further been investigated theoretically and numerically~\cite{elgeti09, elgeti16, daddi18jpcm, wioland16, theers16, degraaf16channel, kuron19}.

Unlike fluid-fluid or fluid-solid interfaces, elastic boundaries generically stand apart because they endow the system with memory.
Such an effect results in a long-lasting anomalous subdiffusive behavior on nearby particles~\cite{daddi16, daddi16b, daddi16c, daddi18epje}.
The emerging subdiffusion can significantly enhance residence time and binding rates and thus may increase the probability to trigger the uptake of particles by living cell membranes via endocytosis~\cite{xiao12, juenger15}. Moreover, theoretical investigations of model microswimmers immersed in an elastic channel have predicted an enhancement in swimming speed as the swimmers deform the flexible boundaries via hydrodynamic flows~\cite{Ledesma:2013}.
In addition, it has been demonstrated that reciprocal motion close to a deformable interface can circumvent the scallop theorem and result to a net propulsion of microswimmers at low Reynolds  numbers~\cite{trouilloud08}.
Theoretically, the motion of a passive particle near a fluid membrane possessing surface tension~\cite{bickel07, bickel14}, bending resistance~\cite{bickel06}, or surface elasticity~\cite{felderhof06, felderhof06b} has thoroughly been studied.
The corresponding diffusion coefficient in the steady limit is found to be universal and identical to that predicted near a hard wall with no-slip boundary conditions~\cite{felderhof06}.

Here, we investigate the influence of nearby elastic boundaries possessing resistance toward shear and bending on the dynamics of microswimmers at low Reynolds number.
Our analytical approach is based on the far-field hydrodynamic multipole representation of active microswimmers and valid in the small-deformation regime.
We find that the shear- and bending-related contributions to the overall induced translational and rotational velocities resulting from the hydrodynamic interactions with an elastic interface may have promotive or suppressive effects.
In the steady limit, the swimming velocities  are found to be independent of the membrane elastic properties and to approach the corresponding values near a no-slip wall.

The remainder of the paper is organized as follows.
In Sec.~\ref{sec:theory}, we present the governing equations of low-Reynolds-number fluid motion and introduce, in the small deformation regime, a relevant model for an elastic interface featuring resistance toward both shear and bending.
In addition, we describe in terms of the multipole expansion of the Stokes equations the self-generated flow field induced by an active microswimmer near an elastic interface.
We then evaluate in Sec.~\ref{sec:swimming} the induced swimming velocities due to hydrodynamic interactions with the interface and discuss the interplay between shear and bending deformation modes, as well as their corresponding roles in the overall dynamics.
Concluding remarks are contained in Sec.~\ref{sec:conclusion}.
\review{Some mathematical details, which are not essential for the understanding of the key messages of our analytical approach, are relegated to the Appendices.}

\section{Theoretical description}
\label{sec:theory}
\begin{figure}
	\centering
	\includegraphics[scale=1]{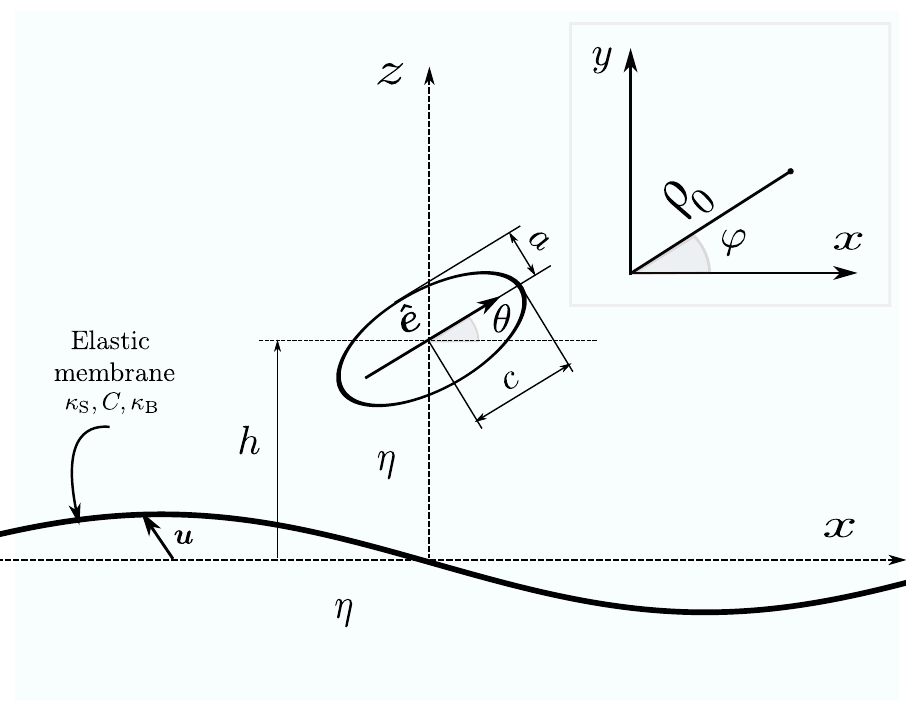}
	\caption{Illustration of the system setup. An axisymmetric active microswimmer modeled as a prolate spheroid is trapped at~$z=h$ above an elastic interface infinitely extended in the \mbox{$xy$-plane}.
	The lengths of the short and long semi-axes are denoted by~$a$ and~$c$, respectively.
	Setting the orientation of the swimmer, the unit vector~$\boldsymbol{\hat{e}}$ points along the symmetry axis of the swimmer.
	The pitch angle of the swimmer relative to the horizontal plane is denoted by~$\theta \in [-\pi/2,\pi/2]$ (complement of the polar angle in spherical coordinates).
	On both sides of the elastic interface, the surrounding fluid is Newtonian and characterized by the same dynamic viscosity~$\eta$.
	The figure shown in the inset is a top view of the local reference frame associated with the microswimmer, where~$\rho_0$ is the radial distance and~$\varphi \in [0,2\pi)$ is the azimuthal orientation.
	}
	\label{illustration-of-system-setup}
\end{figure}

We consider the behavior of an axisymmetric microswimmer near a planar elastic interface of infinite extent in the $xy$-plane, i.e., the $z$-direction is directed normal to that plane. The swimmer is modeled as a prolate spheroid of short semi-axis~$a$ and long semi-axis~$c$, trapped above the elastic interface at position~$z=h$. Here, we adopt a local coordinate system attached to the swimmer such that~$\theta \in [-\pi/2,\pi/2]$ is the pitch angle and~$\varphi \in [0,2\pi)$ is the azimuthal orientation in the $xy$-plane, see Fig.~\ref{illustration-of-system-setup} for a graphical illustration of the system setup.

We model the swimming behavior in the far-field limit (i.e., ~$c \ll h$) by using a combination of fundamental solutions to the Stokes equations in the vicinity of an elastic interface~\cite{chwang75, spagnolie12}. Further details on the swimmer model are provided after stating the exact Green's functions for a point-force singularity near a planar elastic boundary and derivation of the corresponding higher-order singularities that are obtained via a multipole expansion (see Sec.~\ref{sec:swimming}).

\subsection{Low-Reynolds-number hydrodynamics -- Stokes equations}
For a viscous, incompressible Newtonian fluid, the Navier Stokes equations in the overdamped, low-Reynolds-number limit simplify to the time-independent Stokes equations,~\cite{purcell77,elgeti15}
\begin{subequations} \label{stokesEqs}
	\begin{align}
	\eta \bNabla^2 \vect{v}(\R) - \bNabla p(\R) + \vect{f}_\mathrm{B}(\R)   &= 0 \, , \label{stokesEq_1} \\
	\bNabla \cdot \vect{v} (\R) &= 0 \, ,  \label{stokesEq_2}
	\end{align}
\end{subequations}
where $\R$ denotes the spatial coordinate, $\eta$ is the shear viscosity, $\vect{v}$ denotes the fluid velocity, $p$~is the pressure field, and $\vect{f}_\mathrm{B}$~here represents the body force density acting on the fluid domain by the immersed objects.

The fundamental solution of the Stokes equations for a point-force singularity~$\vect{f}_\mathrm{B} = \vect{f} \delta (\R-\R_0)$ (Stokeslet) placed at position~$\R_0$ in an otherwise quiescent unbounded (infinite) fluid domain is expressed in terms of the free-space Green's function given by the Oseen tensor~\cite{happel12,kim13}.
Assuming that the point force is directed along the unit vector~$\e$ such that~$\vect{f} = f \e$, the induced flow and pressure fields reads
\begin{equation}
	\vect{v}_\mathrm{S}^\infty (\R) = \frac{f}{8\pi\eta} \, \bG^\infty (\R, \R_0; \e) \, , \quad
	p_\mathrm{S}^\infty (\R) = \frac{f}{4\pi} \, P^\infty (\R, \R_0; \e) \, ,
\end{equation}
where the Stokeslet solution is given by
	$\bG^\infty (\R, \R_0; \e) =  \left( \e + \left( \e \cdot \vect{\hat{s}} \right) \, \vect{\hat{s}} \right)/s$,
with~$\vect{s} = \R-\R_0$, $s = |\vect{s}|$ denoting the distance from the singularity position, and~$\vect{\hat{s}} = \vect{s}/s$.
Likewise, the corresponding solution for the pressure field is $P^\infty (\R, \R_0; \e) = \e \cdot \vect{\hat{s}}/s^2$.

\subsection{Model for the elastic interface}

The interface is modeled as a~2D elastic sheet made of a hyperelastic material featuring resistance toward both shear and bending.
Shear elasticity of the interface is described by the well-established Skalak model~\cite{skalak73}, which is commonly utilized as a practical model for the description of red blood cell membranes~\cite{Freund_2014,krueger11,krueger12, barthes16}.
The interface resistance toward bending is described by the Helfrich model~\cite{helfrich73,seifert97, guckenberger16, Guckenberger_review}.

For an elastic interface infinitely extended in the $xy$-plane, the linearized tangential and normal traction jumps across the interface due to shear and bending deformation modes are expressed in terms of the displacement field~$\vect{u}$ of the interface relative to the initial planar configuration via~\cite{daddi16}
\begin{subequations}
	\begin{align}
		[\sigma_{zj}] &= -\frac{\kS}{3} \left( \Delta_\parallel u_j + \left(1+2C\right) \partial_j \epsilon \right) \, , \quad j \in \{x,y\} \, , \\
		[\sigma_{zz}] &= \kB \Delta_\parallel^2 u_z \, ,
	\end{align}
\end{subequations}
where~$\kS$ is the shear modulus, $C=\kappa_\mathrm{A}/\kS$ denotes the Skalak parameter (with the area expansion modulus~$\kappa_\mathrm{A}$), and $\kB$ is the bending modulus.
Here we use the notation~$[\sigma_{ij}] = \sigma_{ij}(z=0^+) - \sigma_{ij}(z=0^-)$ to denote the jump in the viscous stress tensor across the elastic interface.
In addition, $\epsilon = \partial_x u_x + \partial_y u_y$ denotes the dilatation function, and~$\Delta_\parallel = \partial_x^2 + \partial_y^2$ stands for the Laplace-Beltrami operator~\cite{deserno15}.
The normal components of the hydrodynamic stress tensor are expressed in the Cartesian coordinate system in the usual way as~$\sigma_{zj} = -p\delta_{zj} + \eta \left( \partial_j v_z + \partial_z v_j \right)$.

To relate the displacement of the elastic interface to the fluid velocity field, we impose a hydrodynamic no-slip boundary condition.
The latter, in Fourier space, takes a particularly simple form in the small-deformation regime.
Specifically~\cite{bickel07}, 
\begin{equation}
	\vect{v} |_{z=0} = i\omega \, \vect{u} \, , \label{no-slip}
\end{equation}
with~$\omega$ being the frequency in the Fourier domain.
Accordingly, the components of the fluid velocity field evaluated at the surface of reference~$z=0$ are assumed to coincide with those of the material points composing the deformable interface.
The particular case of zero frequency corresponds to the \enquote{stick} boundary condition which applies for an infinitely-extended rigid wall~\cite{felderhof06}.
It is worth mentioning that, if the elastic interface undergoes a larger deformation, the no-slip condition stated by Eq.~\eqref{no-slip} takes a nonlinear form because the condition has to be applied at the deformed interface.
This situation has been considered, for instance, in Refs.~\onlinecite{sekimoto93, weekley06, salez15, saintyves16, rallabandi17, daddi18stone, rallabandi18, zaicheng19}. Since our attention here is restricted to the system behavior in the small-deformation regime, for which~$|\vect{u}| \ll h$, applying the no-slip boundary condition at the position of the undisplaced interface is appropriate for our theoretical analysis.

As described in detail in Refs.~\onlinecite{daddi16} and \onlinecite{daddi16c}, the behavior of a particle close to an elastic interface can conveniently be characterized in terms of the two dimensionless parameters
\begin{equation}
	\beta = \frac{6Bh\eta\omega}{\kS} \, ,
	\qquad
	\betaB = 2h \left( \frac{4\eta\omega}{\kB} \right)^{1/3} \, ,
\end{equation}
where~$B=2/(1+C)$.
Note that both~$\beta$ and~$\betaB^3 \propto \omega$, and can thus be viewed as dimensionless frequencies associated with shear and bending deformation modes, respectively.

The exact Green's functions for a point-force singularity acting close to an elastic interface possessing shear and bending rigidities have recently been calculated by some of us, see, e.g., Refs.~\onlinecite{daddi16, daddi18epje} for their details of derivation.
The~$\e$-directed Stokeslet near the elastic interface can be obtained from the tensorial description of the Green's function via
\begin{equation}
	\vect{G} (\R,\R_0; \e) = 8\pi\eta \, \vect{\mathcal{G}} (\R,\R_0) \cdot \e \, . \label{e-directed-Stokeslet}
\end{equation}

The \emph{frequency-dependent} Green's functions~$\vect{\mathcal{G}}$ associated with a point-force exerted at position~$\R_0$ above an elastic interface can be derived using a standard two-dimensional Fourier transform technique~\cite{bickel06, bickel07} and applying the underlying boundary conditions at the planar surface of reference.
\review{Accordingly, the Green's functions can be expressed in terms of convergent infinite integrals over the wavenumber.
Explicit analytical expressions of the components of the Green's functions due to a Stokeslet near an elastic interface are listed for convenience in Appendix~\ref{appendix:greenfunction}.}

\subsection{Multipole expansion}\label{subsec:multipoleExpansion}
The flow field generated by a microswimmer can be decomposed into a multipole expansion of the solution of the Stokes equations [Eq.~\eqref{stokesEqs}] near an elastic interface.
Then, the linearity of the Stokes equations permits the description of the far-field flow induced by a microswimmer in terms of a superposition of different singularity solutions~\cite{spagnolie12}.
While the leading order flow field of a driven particle is a force monopole (Stokeslet) field which decays as $s^{-1}$, force- and torque-free microswimmers typically create a force dipole field in leading order~\cite{lauga09, elgeti15} which decays as $s^{-2}$.
The next-higher-order singularities are the force quadrupole, source dipole, and rotlet dipole,  which all decay as $s^{-3}$.
The Green's functions for higher-order singularities can be obtained as derivatives of the Stokeslet solution~\cite{chwang75}.
For example, for a force dipole (D),
\begin{equation}
		\bG_\mathrm{D} (\R, \R_0; \e, \vect{a}) = \left( \vect{a} \cdot \bNabla_0 \right) \bG(\R, \R_0; \e ) \, ,
		\label{force-dipole-def}
\end{equation}
wherein~$\bNabla_0$ denotes the nabla (gradient) operator taken with respect to the singularity position~$\R_0$.
The force quadrupole (Q) can then be determined from the force dipole as
\begin{equation}
	\bG_\mathrm{Q} (\R, \R_0; \e, \vect{a}, \vect{b}) = \left( \vect{b} \cdot \bNabla_0 \right) \bG_\mathrm{D} (\R, \R_0; \e, \vect{a}) \, .
	\label{force-quadrupole-def}
\end{equation}
In addition, we define the source dipole (SD) singularity which can be derived from a singular potential solution satisfying the Laplace equation~\cite{spagnolie12}.
It can be expressed in terms of the Stokeslet solution via
\begin{equation}
	\bG_\mathrm{SD} (\R, \R_0; \e) = -\frac{1}{2} \, \bNabla_0^2 \bG (\R,\R_0; \e) \, .
	\label{source-dipole-def}
\end{equation}
Further, we define the rotlet dipole (RD) singularity as
\begin{equation}
	\bG_\mathrm{RD} (\R, \R_0; \e, \vect{c}) = \vect{c} \cdot \bNabla_0 \bG_\mathrm{R} (\R, \R_0; \e) \, ,
	\label{rotlet-dipole-def}
\end{equation}
where the Green's function for the rotlet (R) is obtained as
\begin{equation}
  \bG_\mathrm{R}  (\R, \R_0; \e) = \frac{1}{2} \big( \bG_\mathrm{D} (\R, \R_0; \vect{b},\vect{a}) - \bG_\mathrm{D}(\R, \R_0; \vect{a},\vect{b}) \big) \, , \label{rotlet-def}
\end{equation}
where~$\vect{a}$ and~$\vect{b}$ are unit vectors with $\vect{a} \times \vect{b} = \e$ ($\times$ denotes the cross product).
Note that the rotlet is the leading order flow field of a force-free particle but where an external torque is applied.
The flow field due to a rotlet dipole can further be expanded as a combination of two force quadrupoles as
\begin{equation}
  \bG_\mathrm{RD} (\R, \R_0; \e, \vect{c}) = \frac{1}{2} \big( \bG_\mathrm{Q} (\R, \R_0; \vect{b},\vect{a},\vect{c}) - \bG_\mathrm{Q} (\R, \R_0; \vect{a}, \vect{b}, \vect{c}) \big) \, .
  \notag
\end{equation}

\review{Expressions of the higher-order Stokes singularities in an unbounded (infinite) fluid are provided in Appendix~\ref{appendix:higher-order-singularities}.}

In the presence of external forces and torques acting on the microswimmer, the Stokeslet, $\bG (\R, \R_0; \e)$, and rotlet $\bG_\mathrm{R} (\R, \R_0; \e)$, solutions have to be added to our description.
Collecting results, the self-generated flow field induced by an axially-symmetric microswimmer initially located at position~$\R_0$ and oriented along the direction of the unit vector~$\e$ can be written up to third order in inverse distance from the swimmer location as
\begin{equation}
			\vect{v} (\R) =
			\vect{v}_\mathrm{S} (\R) + \vect{v}_\mathrm{R} (\R) + \vect{v}_\mathrm{D} (\R) + \vect{v}_\mathrm{SD} (\R) + \vect{v}_\mathrm{Q} (\R) + \vect{v}_\mathrm{RD} (\R) \, , \label{eq_v}
\end{equation}
where we have defined the velocities
\begin{align*}
                \vect{v}_\mathrm{S} (\R) &= \alpha_\mathrm{S} \, \bG (\e)                  \, , & \vect{v}_\mathrm{R} (\R) &= \alpha_\mathrm{R} \, \bG_\mathrm{R} (\e),      \\        
  		\vect{v}_\mathrm{D} (\R) &= \alpha_\mathrm{D} \, \bG_\mathrm{D}(\e, \e) \, , & \vect{v}_\mathrm{SD} (\R) &= \alpha_\mathrm{SD} \, \bG_\mathrm{SD}(\e) \, ,  \\
		\vect{v}_\mathrm{Q} (\R) &= \alpha_\mathrm{Q} \, \bG_\mathrm{Q}(\e, \e, \e) \, , & \vect{v}_\mathrm{RD} (\R) &= \alpha_\mathrm{RD} \, \bG_\mathrm{RD} (\e, \e) \, , 
	\end{align*}
not writing the dependence of the flow singularities on~$\R$ and~$\R_0$ explicitly any longer.

The Stokeslet coefficient~$\alpha_\mathrm{S}$ has dimension of (length)$^2$(time)$^{-1}$, the rotlet coefficient~$\alpha_\mathrm{R}$ and dipolar coefficient~$\alpha_\mathrm{D}$ have dimension of (length)$^3$(time)$^{-1}$, whereas the remaining higher-order multipole coefficients~$\alpha_\mathrm{SD}$, $\alpha_\mathrm{Q}$, and~$\alpha_\mathrm{RD}$ have dimensions of (length)$^4$(time)$^{-1}$.
The magnitude and sign of these coefficients depend on the propulsion mechanism as well as on the swimmer shape.
For a valuable discussion on the physical meaning and interpretation of these singularities, we refer the reader to recent works by Spagnolie and Lauga~\cite{spagnolie12}, and Mathijssen~\textit{et al.}~\cite{mathijssen16jfm}.

\section{Swimming near an elastic interface}
\label{sec:swimming}
In the presence of confining boundaries, the swimming direction $\e$ of the microswimmer and its distance $h$ from the boundary dictate the hydrodynamic flows, as sketched in Fig.~\ref{illustration-of-system-setup}. The orientation $\e$ is described by the unit vector
\begin{equation}
	\e = \left( \cos\theta \cos\varphi, \cos\theta\sin\varphi, \sin\theta \right) \, ,
	\label{Orientierung}
\end{equation}
where, again, $\theta$ denotes the pitch angle (such that~$\theta=0$ corresponds to a swimmer that is aligned parallel to the interface), and~$\varphi$ is the azimuthal orientation that we, without loss of generality, set initially to zero.

The total self-generated flow field of the swimmer expressed by Eq.~\eqref{eq_v} can be decomposed into terms of the bulk contribution~$\vect{v}^\infty$ and a correction~$\vect{v}^*$ that is required to satisfy the boundary conditions at the elastic interface,
\begin{equation}
	\vect{v} = \vect{v}^\infty + \vect{v}^* \, .
\end{equation}
The latter encompasses the Stokeslet contribution to the flow field that we have determined in previous works~\cite{daddi16,daddi17} in addition to the higher-order singularity solutions that we calculate here.
It is worth emphasizing that~$\vect{v}_\infty$ is the sum of the bulk flow fields of the different multipoles such that
\begin{equation}
	\vect{v}^\infty = \lim_{\beta, \betaB\to\infty} \vect{v} \, .  
\end{equation}

The induced translational and rotational velocities due to the fluid-mediated hydrodynamic interactions between the elastic interface and a microswimmer of prolate ellipsoidal shape located at position~$\R_0$ are provided by Fax\'en's laws~\cite{kim13} as
\begin{subequations} \label{eq-HI}
	\begin{align}
		\vect{v}^\mathrm{HI} &= \left. \vect{v}^* (\R) \right|_{\R = \R_0} \, , \label{v-HI} \\
		\boldsymbol{\Omega}^\mathrm{HI} &= \left. \frac{1}{2} \bNabla \times \vect{v}^* (\R)
		+ \Gamma \e \times \left( \vect{E}^*(\R) \cdot \e \right) \right|_{\R = \R_0} \, . \label{Omega-HI}
	\end{align}
\end{subequations}
These expressions have been restricted to leading order in swimmer length~$c$.
Here, $\vect{E}^* = \left( \bNabla \vect{v}^* + \left(\bNabla \vect{v}^*\right)^\mathrm{T} \right)/2$ is the rate-of-strain tensor associated with the reflected flow, with~T denoting the transpose.  Further, $\Gamma = (\gamma^2-1)/(\gamma^2+1) \in [0,1)$ is a shape factor (also known as Bretherton constant~\cite{ryan11, clement16}) that depends on the aspect ratio $\gamma$ of the prolate spheroidal microswimmer, defined as the ratio of major to minor semi-axes, i.e.\ $\gamma = c/a \ge 1$.
It vanishes for a sphere and approaches one for needle-like particles of large aspect ratio.
Higher-order correction terms in~$\Gamma$ to the induced hydrodynamic fields can be obtained using the multipole method, see, e.g., Ref.~\onlinecite{daddi17}.

Due to the linearity of the Stokes equations [Eqs.~\eqref{stokesEqs}] we can consider the effect of each higher-order singularity on the swimming behavior independently. Thus, in the following we provide solutions for the translational and rotational velocities, $\vect{v}^\mathrm{HI}$ and $\boldsymbol{\Omega}^\mathrm{HI}$, induced by fluid-mediated hydrodynamic couplings of the individual contributions with the nearby elastic boundary.

Remarkably, the total velocities due to hydrodynamic interactions with an elastic interface endowed simultaneously with both shear and bending resistances can be written as a superposition of the velocities induced by hydrodynamic interactions with an interface of pure shear $(\betaB\to\infty)$ and pure bending $(\beta\to\infty)$ resistances.
Accordingly, the total wall-induced linear and angular velocities can be obtained by evaluating both contributions independently,
\begin{subequations}
	\begin{align}
		\vect{v}^\mathrm{HI}            &= \left.  \vect{v}^\mathrm{HI} \right|_\mathrm{S}
		                     \,+ \left.  \vect{v}^\mathrm{HI} \right|_\mathrm{B} \, ,  \\
		\boldsymbol{\Omega}^\mathrm{HI} &= \left.  \boldsymbol{\Omega}^\mathrm{HI} \right|_\mathrm{S}
				                     + \left.  \boldsymbol{\Omega}^\mathrm{HI} \right|_\mathrm{B} \, ,
	\end{align}
\end{subequations}
where the subscripts~S and~B stand for shear and bending, respectively. 
However, it is worth mentioning that this is only true for a planar elastic interface.
For curved interfaces, a coupling between shear and bending deformation modes exists~\cite{kovsmrlj17, daddi17b, daddi17c, daddi18creeping, hoell19creeping}.

Near a no-slip wall, the induced hydrodynamic interactions of the multipole flow fields created by a microswimmer located at a given position and orientation is independent of time~\cite{spagnolie12} (assuming that the strengths of the singularities are constant).
This is in contrast to an elastic interface where memory effects can lead to time-dependent contributions  $\vect{v}^\mathrm{HI}(t)$ and $\boldsymbol{\Omega}^\mathrm{HI}(t)$.
One way to realize such a time dependence is to assume that the microswimmer is initially at rest with a given orientation $(\theta, \varphi)$ at a distance $h$ from the interface and suddenly starts to swim and sets  the surrounding fluid into motion at time~$t=0$. 
However, we do not allow the microswimmer to actually \textit{move} towards the interface but its position and orientation is kept fixed by applying just the right external forces~$\vect{F}^\mathrm{ext}$ and torques~$\vect{T}^\mathrm{ext}$, e.g., via optical traps, aligning magnetic fields, or other micro-manipulation techniques.
Denoting by~$v_0$ the bulk swimming speed, i.e., in the absence of the confining interface,
the swimming velocities and rotation rates are related to the external forces and torques required to trap the swimmer near the interface via
\begin{equation}
	\begin{pmatrix}
		 v_0 \e + \vect{v}^\mathrm{HI} (t) \\
		\boldsymbol{\Omega}^\mathrm{HI} (t)
	\end{pmatrix} + \boldsymbol{\mu} \cdot
	\begin{pmatrix}
		\vect{F}^\mathrm{ext} (t) \\
		\vect{T}^\mathrm{ext} (t)
 	\end{pmatrix} = \vect{0} \, ,
\end{equation}
Note, the forces and torques are zero for $t<0$, but finite and time-dependent for $t \ge 0$ when the flow fields created by the microswimmers interact with the elastic interface. Here $\boldsymbol{\mu}$ is the position- and orientation-dependent hydrodynamic grand mobility tensor of a spheroid near an elastic interface~\cite{daddi17}.
We have neglected thermal fluctuations and all possible steric interactions with the interface.
We were able to calculate  $\vect{v}^\mathrm{HI} (t)$ and $\boldsymbol{\Omega}^\mathrm{HI} (t)$ for all considered multipole flows. The solutions for $\vect{v}^\mathrm{HI} (t)$ are shown in Tab.~\ref{tab:time-dependent-expressions} and Tab.~\ref{tab:coeff}. Similar expressions exist for $\boldsymbol{\Omega}^\mathrm{HI} (t)$ but they are not shown here because of their complexity and lengthiness.

In the following we discuss the different contributions stemming from the different multipoles.
Before doing so, we present typical numbers which we used to produce the results shown below.
The shear and bending properties of the elastic surface entail a characteristic time scale of shear as~$T_\mathrm{S} = 6\eta h/(B\kS)$, in addition to a characteristic time scale of bending as~$T_\mathrm{B} = 8\eta h^3/\kB$~\cite{daddi16}. Thus, we define the scaled times~$\tauS = t/T_\mathrm{S}$ and~$\tauB = t/T_\mathrm{B}$ associated with shear and bending deformation modes, respectively. 
Note that, for~$h = \big( 3\kB/(4B\kS) \big)^{1/2}$, it follows that~$T_\mathrm{S} = T_\mathrm{B}$.
This corresponds to the situation in which both shear and bending equally manifest themselves in the system at intermediate time scales~\cite{daddi16b}.
In typical situations~\cite{Freund_2014}, elastic red blood cells have a shear modulus $\kS = 5\times 10^{-6}$~N/m, a Skalak ratio~$C=100$, and a bending modulus $\kB=2\times 10^{-19}$~Nm.
By considering a dynamic viscosity of the surrounding Newtonian viscous fluid~$\eta = 1.2 \times 10^{-3}$~Pa\,$\cdot$\,s, as well as a micron-sized swimmer of size~$a = 10^{-6}$~m located above the interface at~$h = 5a$, it follows that $T_\mathrm{S} \simeq 0.36$~s and~$T_\mathrm{B} = 6$~s.
Therefore, at later times, bending effects are expected to manifest themselves in a more pronounced way than shear.
For the results presented below, we use $\tau := \tauS = 16\tauB$ as the scaled time of the system. 

We distinguish contributions relevant for force- and torque-free swimming and contributions stemming from external forcing, where particular focus lies on a trapped microswimmer in the vicinity of an elastic interface.

%%%%% Force Dipole

\subsection{Force- and torque-free contribution}
Here we discuss the swimming behavior of an active agent near an elastic boundary by following the theoretical framework discussed in Sec.~\ref{sec:theory}. We consider different higher-order singularities that describe features of the swimming motion of a variety of active agents. In addition to the leading-order far-field of a microswimmer in terms of a force-dipole ($1/s^2$), we consider further details of the propulsion mechanisms that contribute to the flow field with the order of $1/s^3$. These include, for example, contributions of the finite size cell body, the anisotropy in the swimming mechanism, and the rotation/counter-rotation of body parts during swimming. Yet, the importance of the contribution of each of these singularities depends strongly on the geometry of the active agent, its swimming mechanism, and its distance from the elastic interface.

\subsubsection{Force dipole}
The flow field induced by a force dipole, $\vect{v}_\mathrm{D} (\R) = \alpha_\mathrm{D} \, \bG_\mathrm{D}(\e, \e)$, is the leading contribution to describe the hydrodynamics of many microswimmers, which are net-force-free by definition \cite{purcell77}.
The sign of the dipolar coefficient~$\alpha_\mathrm{D}$ distinguishes between \emph{pusher} ($\alpha_\mathrm{D}>0$) and \emph{puller} ($\alpha_\mathrm{D}<0$) microswimmers. Some bacterial microorganisms, such as \textit{E. coli}, exploit (bundles of) helical filaments called flagella for their propulsion, the rotation of which causes the entire bacterium to move forward in a corkscrew-like motion~\cite{berg08,chevance08,lisicki19}.
Here, the translation-rotation coupling of the hydrodynamic friction of the flagellum yields a net propulsion of the swimmer.
Since these swimmers push out the fluid along their swimming axis, they are referred to as pushers. 
Another broad class of microswimmers, including, for example, the algae \textit{Chlamydomonas reinhardtii} \cite{drescher10}, pull in (averaged over one whole swimming stroke) the fluid along the axis parallel to their swimming direction, and are thus classified as pullers.

Both pushers and pullers may conveniently be modeled, e.g., via minimal models based on the insertion of force centers that co-move with the body of the swimmer \cite{menzel16,hoell17,adhyapak17,daddi18prf,schwarzendahl18}, or as squirmers~\cite{downton09,pedley16, pedley16, zottl18}.
The latter are driven by prescribed tangential velocities at their (spherical or ellipsoidal) surfaces and were introduced to model microorganisms that self-propel by the beating of cilia covering their bodies~\cite{lighthill52, blake71cilia, brennen77, pak14}.
The squirmer model has been previously used to address, e.g., the hydrodynamic interaction between two swimmers~\cite{ishikawa07, gotze10}, the influence of an imposed external flow field on the swimming behavior~\cite{zottl12, zottl13}, or low-Reynolds-number locomotion in complex fluids~\cite{datt15, nganguia17, nganguia18, pietrzyk19}.

We now return to the mathematical problem and remark that a tilted force dipole (that is directed along~$\e$) can be expressed in terms of force dipoles aligned parallel and perpendicular to the elastic interface as~\cite{lopez14}
\begin{equation}
 \begin{split}
  \bG_\mathrm{D} (\e,\e) &= \bG_\mathrm{D}(\e_x,\e_x) \cos^2\theta
  + \bG_\mathrm{D}(\e_z,\e_z) \sin^2 \theta \\
  &\quad
  + \bG_\mathrm{SS} (\e_x,\e_z) \sin (2\theta) \, ,
 \end{split}
\end{equation}
where~$\bG_\mathrm{SS}$ is the symmetric part of the force dipole, commonly referred to as stresslet,
$\bG_\mathrm{SS} (\vect{a}, \vect{b}) = \left( \bG_\mathrm{D}(\vect{b}, \vect{a}) + \bG_\mathrm{D}(\vect{a}, \vect{b}) \right)/2$.
By inserting the Stokeslet solution (see \review{Appendix}~\ref{appendix:greenfunction}) into Eq.~\eqref{force-dipole-def}, the self-generated dipolar flow field~$\vect{v}_\mathrm{D} (\R)$ can be evaluated and expressed in terms of infinite integrals over the wavenumber. 
\review{The frequency-dependent components of the induced translational, $\vect{v}^\mathrm{HI}$, and rotational, $\boldsymbol{\Omega}^\mathrm{HI}$, velocities, of the microswimmer resulting from dipolar interactions with the elastic interface, as given by Eq.~\eqref{eq-HI}, are listed in integral form in Tab.~\ref{tab:freq-dependent-expressions} of Appendix~\ref{appendix:tables}.}
The velocities in Fourier space depend on the dipolar coefficient~$\alpha_\mathrm{D}$, the distance~$h$ from the elastic interface, the orientation~$\theta$ of the swimmer with respect to the interface, as well as on the dimensionless frequencies~$\beta$ and $\betaB$, reflecting shear and bending contributions, respectively.

\begin{figure*}[htp]
	% \scalebox{0.66}{\input{swimmingVelo-Higher-Order}}
	\includegraphics[scale=1]{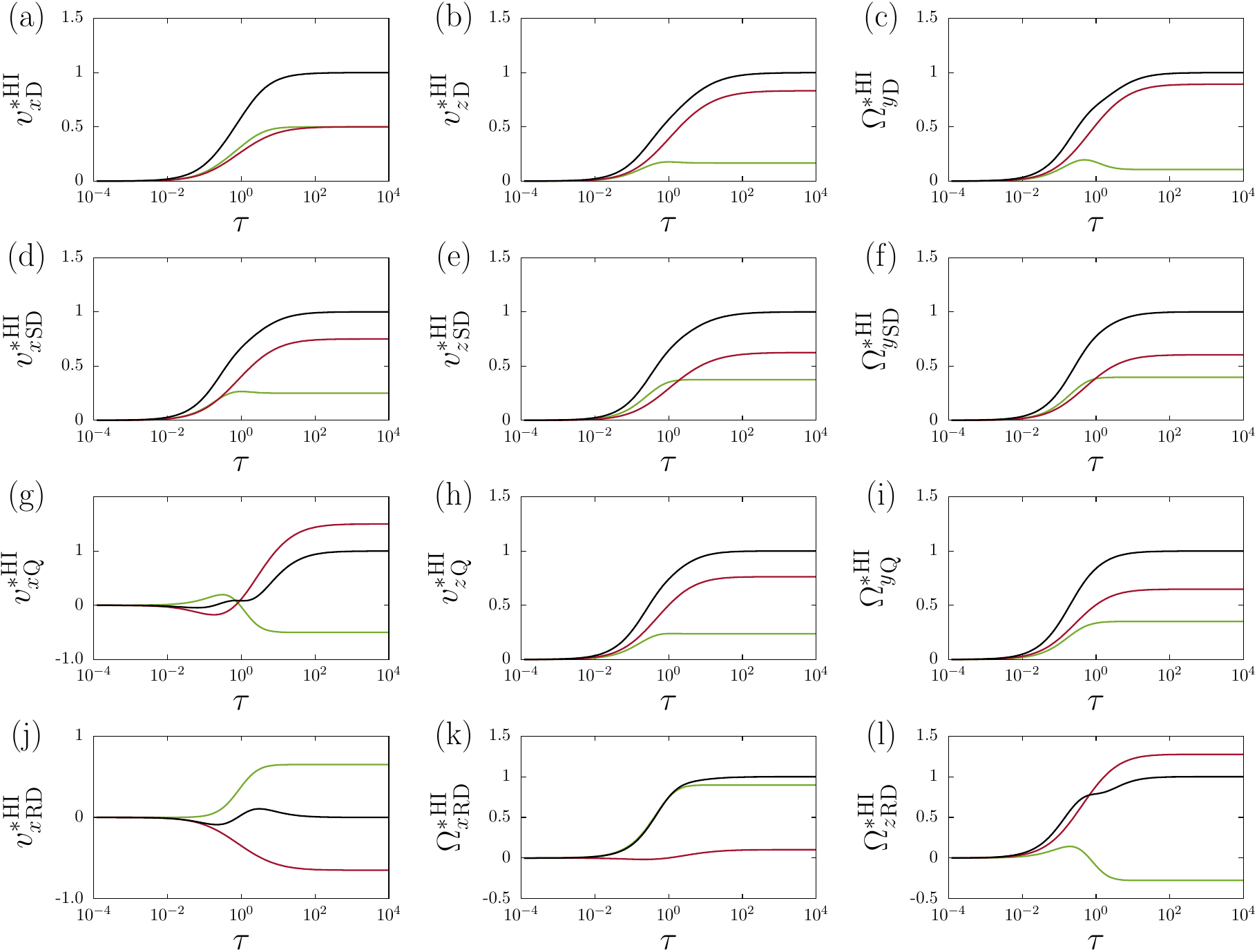}
	\caption{(Color online)
	Evolution of the scaled induced translational and rotational swimming velocities associated with a force dipole [$(a)$ -- $(c)$], source dipole [$(d)$ -- $(f)$], force quadrupole [$(g)$ -- $(i)$], and rotlet dipole [$(j)$ -- $(l)$], resulting from hydrodynamic interactions with a planar elastic interface of pure shear (green), pure bending (red), or both shear and bending (black) resistances.
	The swimmer has an aspect ratio~$\gamma = 4$ and is oriented by a pitch angle $\theta = \pi/6$ relative to the horizontal direction.
	Here, the velocities are scaled by the corresponding hard-wall limits listed in Tab.~\ref{table_vel_wall}, except that the $x$-component of the rotlet dipolar contribution shown in panel~$(j)$ is scaled by~$\alpha_\mathrm{RD}/(8h^4)$ (because this component vanishes in the steady limit).
	The scaled time is~$\tau := \tauS = 16\tauB $.
  \label{fig:swimmingvel}}
\end{figure*}

\begin{table*}
\begin{minipage}{\textwidth}
	\hspace{-18.cm}
	\vspace{-2.3cm} {\Large (a)}
\end{minipage}
\begin{minipage}{\textwidth}
	\hspace{-18.cm}
	\vspace{-19.2cm} {\Large (b)}
\end{minipage}
\begin{center}
%	\begin{table}
	\def\arraystretch{2}%
		\begin{tabular}{|cc|c|c|c|}
				\cline{2-5}
				\multicolumn{1}{c}{} & \multicolumn{1}{|c|}{\makecell{~Interface type~~}} & $h^n v_x^\mathrm{HI}$ & $h^n  v_z^\mathrm{HI}$ & $h^{n+1}  \Omega_y^\mathrm{HI}$ \\
				\cline{2-5}
				\multicolumn{1}{c}{}\\[-20pt]\hline
				\multirow{3}{*}{\rotatebox[origin=c]{90}{\makecell{~ \\ Force dipole \\ ~}}} & \multicolumn{1}{|c|}{Shear}
				& {\small $ \frac{3 \alpha_\mathrm{D}}{16 } \, \sin (2\theta)$}
				& {\small $-\frac{\alpha_\mathrm{D} }{16 } \left( 3\cos^2\theta-2 \right)$}
				& {\small $\frac{3\alpha_\mathrm{D}}{64  } \, \Gamma  \sin (2\theta) \cos^2 \theta$} \\
				\cline{2-5}
				& \multicolumn{1}{|c|}{Bending}
				& {\small $ \frac{3\alpha_\mathrm{D}}{16 } \, \sin (2\theta)$}
				& {\small $-\frac{5\alpha_\mathrm{D} }{16 } \, (3\cos^2\theta-2)$}
				& {\small $\frac{3\alpha_\mathrm{D} }{64 } \, \sin (2\theta) \left(4+\Gamma \left( 4-3\cos^2\theta\right) \right)$}  \\
				\cline{2-5}
				& \multicolumn{1}{|c|}{Hard wall}
				& {\small $ \frac{3\alpha_\mathrm{D} }{8  } \, \sin (2\theta)$}
				& {\small $-\frac{3 \alpha_\mathrm{D}}{8 } \, \left(3\cos^2\theta-2 \right)$}
				& {\small $\frac{3 \alpha_\mathrm{D}}{32 } \, \sin (2\theta) \left( 2+\Gamma(2-\cos^2\theta) \right)$} \\
				\hline\hline
				\multirow{3}{*}{\rotatebox[origin=c]{90}{Source dipole}} & \multicolumn{1}{|c|}{Shear}
				& {\small $-\frac{\alpha_\mathrm{SD}}{16 } \, \cos\theta$}
				& {\small $-\frac{3\alpha_\mathrm{SD} }{8 } \, \sin\theta$}
				& {\small $-\frac{3\alpha_\mathrm{SD} }{16 } \, \cos\theta
				\left( 1+\Gamma\left(2-\cos^2\theta\right) \right)$} \\
				\cline{2-5}
				& \multicolumn{1}{|c|}{Bending}
				& {\small $-\frac{3\alpha_\mathrm{SD}}{16 } \, \cos\theta$}
				& {\small $-\frac{5 \alpha_\mathrm{SD}}{8 } \, \sin\theta $}
				& {\small $-\frac{3\alpha_\mathrm{SD}}{16 } \, \cos\theta \left( 1+2\Gamma\left(2-\cos^2\theta\right) \right)$} \\
				\cline{2-5}
				& \multicolumn{1}{|c|}{Hard wall}
				& {\small $-\frac{\alpha_\mathrm{SD}}{4 } \, \cos\theta$}
				& {\small $- \alpha_\mathrm{SD} \sin\theta $}
				& {\small $-\frac{3\alpha_\mathrm{SD}}{16 } \, \cos\theta \left( 2+3\Gamma(2-\cos^2\theta) \right) $} \\
				\hline\hline
				\multirow{3}{*}{\rotatebox[origin=c]{90}{Quadrupole}} & \multicolumn{1}{|c|}{Shear}
				& {\small $\frac{\alpha_\mathrm{Q}}{32 } \, \cos\theta \left( 21\cos^2\theta-16 \right)$}
				& {\small $\frac{3\alpha_\mathrm{Q}}{8 } \, \sin\theta \cos^2\theta$}
				& {\small $\frac{3\alpha_\mathrm{Q}}{64 } \, \cos\theta
				\left( 3\Gamma\cos^4\theta+2(1-2\Gamma)\cos^2\theta+8\Gamma \right)$} \\
				\cline{2-5}
				& \multicolumn{1}{|c|}{Bending}
				& {\small $\frac{3\alpha_\mathrm{Q}}{32 } \, \cos\theta \left( 11\cos^2\theta-8 \right)$}
				& {\small $\frac{\alpha_\mathrm{Q}}{8 } \, \sin\theta \left( 15\cos^2\theta-4 \right)$}
				& {\small $\frac{3\alpha_\mathrm{Q}}{64 } \, \cos\theta \left( -9\Gamma\cos^4\theta+2(11+8\Gamma)\cos^2\theta+8(\Gamma-2) \right)$} \\
				\cline{2-5}
				& \multicolumn{1}{|c|}{Hard wall}
				& {\small $\frac{\alpha_\mathrm{Q}}{16 } \, \cos\theta \left( 27\cos^2\theta-20 \right)$}
				& {\small $\frac{\alpha_\mathrm{Q}}{4 } \, \sin\theta \left( 9\cos^2\theta-2 \right) $}
				& {\small $\frac{3\alpha_\mathrm{Q}}{32 } \, \cos\theta \left( -3\Gamma\cos^4\theta+6(\Gamma+2)\cos^2\theta+8(\Gamma-1) \right) $} \\
				\hline
		\end{tabular}
		\\[10pt]
		\begin{tabular}{|cc|c|c|c|}
		 	\cline{2-5}
			\multicolumn{1}{c}{} & \multicolumn{1}{|c|}{~Interface type~~}
			& $h^3 v_y^\mathrm{HI}$~~ & $h^{4} \Omega_x^\mathrm{HI}$ & $h^{4} \Omega_z^\mathrm{HI}$ \\
			\cline{2-5}
			\multicolumn{1}{c}{}\\[-20pt]\hline
			\multirow{3}{*}{\rotatebox[origin=c]{90}{\makecell{~ \\ Rotlet dipole \\ ~}}}
			& \multicolumn{1}{|c|}{Shear}
			& {\small $\frac{3\alpha_\mathrm{RD}}{32 } \, \sin (2\theta)$}
			& {\small $\frac{3\alpha_\mathrm{RD}}{16 } \, \sin (2\theta)$}
			& {\small $-\frac{3\alpha_\mathrm{RD}}{32 } \left(3\cos^2\theta-2 \right)$} \\
			\cline{2-5}
			& \multicolumn{1}{|c|}{Bending}
			& {\small $-\frac{3\alpha_\mathrm{RD}}{32 } \, \sin (2\theta)$}
			& ~~~~{\small $\frac{3\alpha_\mathrm{RD}}{64 } \, \sin (2\theta) \left( 2 + \Gamma\left( 3\cos^2\theta -4\right) \right)$}~~~~
			& {\small $\frac{3\alpha_\mathrm{RD}}{32 } \, \Gamma  \cos^2\theta  \left(4-3\cos^2\theta \right)$} \\
			\cline{2-5}
			& \multicolumn{1}{|c|}{Hard wall}
			& 0
			& {\small $\frac{3\alpha_\mathrm{RD}}{64 } \, \sin (2\theta) \left( 6+\Gamma(3\cos^2\theta-4) \right)$}
			& ~~~{\small $-\frac{3\alpha_\mathrm{RD}}{32 } \left( 3\Gamma\cos^4\theta+(3-4\Gamma)\cos^2\theta-2 \right) $}~~~ \\
			\hline
		\end{tabular}
		\caption{
		Expressions of the induced translational and rotational swimming velocities resulting from~$(a)$ force dipolar, source dipolar, force quadrupolar, and~$(b)$ rotlet dipolar hydrodynamic interactions with a planar elastic interface in the steady limit.
		Here, $n = 2$ for the force dipole and $n=3$ for the source dipole and force quadrupole.
		The swimming velocities near a no-slip hard wall are obtained by linear superposition of the shear- and bending-related contributions.
    \label{table_vel_wall}
		}
%	\end{table}
\end{center}
\end{table*}

\begin{figure*}
	% \scalebox{0.66}{\input{swimmingVelo-StokesletUndRotlet}}
	\includegraphics[scale=1]{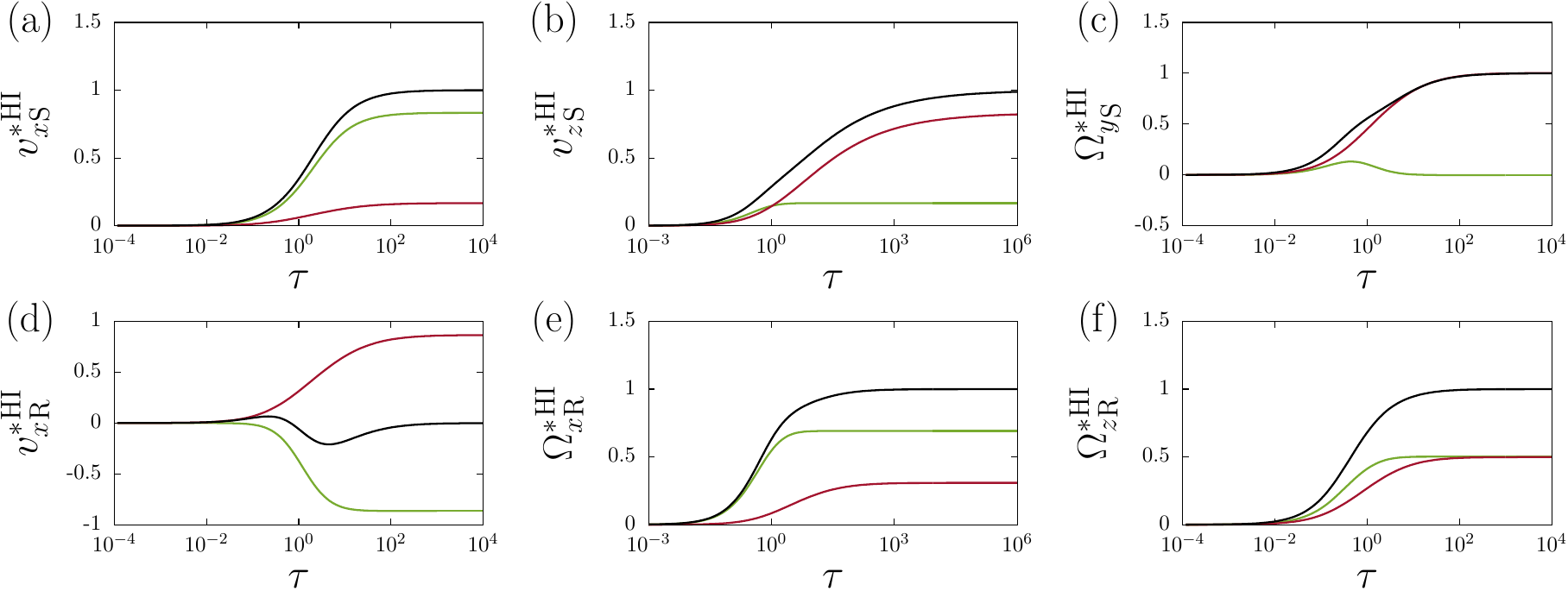}
	\caption{(Color online)
	Evolution of the scaled swimming velocities associated with a Stokeslet [$(a)$ -- $(c)$] and rotlet [$(d)$ -- $(f)$] due to hydrodynamic interactions with an elastic interface showing pure shear (green), pure bending (red), or both shear and bending rigidities (black).
	Here, the swimmer has an aspect ratio~$\gamma = 4$ and an orientation~$\theta = \pi/6$ with respect to the horizontal direction.
	The velocities are scaled by the corresponding hard-wall values except that the $x$~component of the rotlet contribution is scaled by~$\alpha_\mathrm{R}/(8h^4)$.
	We set~$\tau := \tauS = 16\tauB $.
  \label{fig:swimmingvel_stokeslet_rotlet}
	}
\end{figure*}

In Fig.~\ref{fig:swimmingvel}~$(a)$ -- $(c)$, we present the time evolution of the induced swimming velocities and rotation rates due to dipolar hydrodynamic interactions with a planar elastic interface.
The latter has only energetic resistance toward shear (green), only energetic resistance toward bending (red), or simultaneously possesses both shear and bending resistances (black).
Here, we consider a spheroidal swimmer with an aspect ratio~$\gamma=4$ (corresponding to a shape factor~$\Gamma = 15/17$), as measured experimentally for the bacterium \textit{Bacillus subtilis}~\cite{ryan13}.
The swimmer is inclined by a pitch angle $\theta=\pi/6$ with respect to the horizontal direction.
Results are rendered dimensionless by scaling with the corresponding hard wall limits listed in Tab.~\ref{table_vel_wall}.
As already mentioned, the total swimming velocity near a planar interface with both shear and bending resistance is obtained by linearly superimposing the individual contributions stemming from each deformation mode.

The translational and rotational velocities of the microswimmer induced by the presence of the elastic interface amount to small values at short times $(\tau \ll 1)$, because the interface is still relatively undeformed and therefore hardly imposes any elastic resistance toward the flow field induced by the microswimmer.
Consequently, the system exhibits initially a \enquote{bulk-like} behavior.
For increasing times, such that~$\tau \simeq 1$, the presence of the elastic interface becomes more noticeable.
The induced swimming velocities monotonically increase in magnitude before reaching at long times $(\tau \gg 1)$ the steady limits.
These correspond to the velocities induced near a no-slip wall and are independent of the membrane shear and bending properties. Therefore, the elasticity of the boundary only contributes at intermediate time scales to the temporal changes of the swimming behavior, whereas, in the steady state, the swimmer essentially experiences the response of the fully deformed interface that does not change its overall shape of deformation any longer.
It is worth emphasizing that the hard-wall limits are reached (if and) only if the interface is simultaneously endowed with resistance toward shear and bending.
Interestingly, at intermediate time scales, the shear-related contribution to the rotational velocity~[Fig.~\ref{fig:swimmingvel}~$(c)$] exceeds to a certain extent its steady value.

In the steady limit, the sign and magnitude of the swimming velocities are strongly dependent on the dipolar coefficient~$\alpha_\mathrm{D}$ as well as on the pitch angle~$\theta$.
In this situation, because~${v_z}_\mathrm{D}^\mathrm{HI} \propto -\alpha_\mathrm{D} \left(3\cos^2\theta-2\right)$ for all interface types (see Tab.~\ref{table_vel_wall}), it follows that, for a small pitch angle, such that $|\theta| < \arccos \left(\sqrt{6}/3\right)$, a pusher-type microswimmer~$(\alpha_\mathrm{D}>0)$ tends to be attracted toward the interface, while a puller~$(\alpha_\mathrm{D}<0)$ tends to be repelled away from it.
This behavior is purely hydrodynamic in origin as has been discussed earlier by Lauga and collaborators for the case of a hard wall~\cite{berke08,spagnolie12}.
In particular, the hard-wall limits are predominately determined by the bending-related contribution.
This implies that, for the dipolar hydrodynamic interactions, the effect due to the bending rigidity is more pronounced than that due to shear.
In addition, since~${\Omega_y}_\mathrm{D}^\mathrm{HI} \propto \alpha_\mathrm{D} \sin \left(2\theta\right)$, a pusher-type swimmer tends to be oriented along the parallel direction~($\theta=0$ is a stable fixed point), while the interface tends to align a puller in the direction normal to the interface~$(\theta = \pm \pi/2)$. 
Hence, in the absence of external trapping, a puller will tend to swim either toward or away from the interface, depending on whether it is initially pitched toward $(\theta <0)$ or away from the interface $(\theta >0)$. 
Particularly, the extensional flow and the shear-related contribution to the rotation rate vanishes for a sphere~$(\Gamma=0)$.
In such a case, the reorientation of the swimmer is solely dictated by the interface bending properties.

In addition to the leading-order contribution of a force dipole, next-higher-order singularity solutions are useful to describe details of the propulsion mechanism of an active agent. The time-dependent translational and rotational velocities induced by higher-order singularities close to the elastic surface for the start-up motion from static condition are presented \review{in Tab.~\ref{tab:time-dependent-expressions} of Appendix~\ref{appendix:tables},} and the steady limits are shown in Tab.~\ref{table_vel_wall}.

\subsubsection{Source dipole}

The far-field hydrodynamic flows induced by the finite size of a swimming object can be described by a \emph{source dipole}, $\vect{v}_\mathrm{SD} (\R) = \alpha_\mathrm{SD} \bG_\mathrm{SD}(\e)$.
For the type of microswimmers that propel themselves by means of activity on their surfaces, as it is the case for many active colloidal particles~\cite{howse07, liebchen16, LiebchenUndLoewenHabenImmerRecht} or ciliated microorganisms~\cite{stone96, sareh13}, a source dipolar coefficient~$\alpha_\mathrm{SD}>0$ is expected. 
In contrast to that, it is expected that~$\alpha_\mathrm{SD}<0$ for non-ciliated but flagellated microswimmers~\cite{mathijssen16jfm}.

We now consider the scenario of a microswimmer initially at rest before starting to pump the fluid, in a way analogous to what we have introduced in the previous discussion regarding the force dipole contribution.
The respective scaled induced translational and rotational velocities resulting from source dipolar hydrodynamic interactions exhibits a similar logistic sigmoid curve varying between 0 and 1, see Fig.~\ref{fig:swimmingvel} $(d)$ -- $(f)$. 
Similar as for the force dipole contribution, at long times the corresponding values of a no-slip wall are approached.
The bending-related contribution to the swimming velocities is found to be once again more pronounced than that due to shear resistance.

For all types of interface, the induced normal swimming velocity in the steady limit can be cast into the form ${v_z}_\mathrm{SD}^\mathrm{HI} \propto -\alpha_\mathrm{SD} \sin\theta$.
Therefore, the swimmer tends to be attracted to the interface for~$\alpha_\mathrm{SD}>0$ when it is oriented toward it $(\theta<0)$ and tends to be repelled from the interface otherwise.
Moreover, since ${\Omega_y}_\mathrm{SD}^\mathrm{HI} \propto -\alpha_\mathrm{SD} \cos\theta$ it follows that $\theta = \pi/2$ is a stable fixed point for $\alpha_\mathrm{SD} > 0$, thus favoring the escape of the swimmer from the interface in the absence of external trapping.
In contrast to that, $\theta = -\pi/2$ is a stable fixed point for~$\alpha_\mathrm{SD} < 0$, leading to hydrodynamic trapping of the swimmer near the interface.

\begin{table*}
\begin{minipage}{\textwidth}
	\hspace{-16.4cm}
	\vspace{-2.4cm} {\Large (a)}
\end{minipage}
\begin{minipage}{\textwidth}
	\hspace{-16.4cm}
	\vspace{-8.8cm} {\Large (b)}
\end{minipage}
\begin{center}
%	\begin{table}
	\def\arraystretch{2}%
		\begin{tabular}{|cc|c|c|c|}
				\cline{2-5}
				\multicolumn{1}{c}{} & \multicolumn{1}{|c|}{\makecell{~Interface type~~}} & $h v_x^\mathrm{HI}$ & $h  v_z^\mathrm{HI}$ & $h^2  \Omega_y^\mathrm{HI}$ \\
				\cline{2-5}
				\multicolumn{1}{c}{}\\[-20pt]\hline
				\multirow{3}{*}{\rotatebox[origin=c]{90}{\makecell{~ \\ Stokeslet \\ ~}}} & \multicolumn{1}{|c|}{Shear}
				& $-\frac{5 \alpha_\mathrm{S}}{8} \, \cos\theta $
				& ~~~~~~$ -\frac{\alpha_\mathrm{S}}{4} \, \sin\theta $~~~~~~
				& $ \frac{\alpha_\mathrm{S}}{16} \, \cos\theta \left(2-3\Gamma \cos^2\theta\right)$ \\
				\cline{2-5}
				& \multicolumn{1}{|c|}{Bending}
				& $-\frac{\alpha_\mathrm{S}}{8} \, \cos\theta $
				& $ -\frac{5 \alpha_\mathrm{S}}{4} \, \sin\theta $
				& ~~~~~~~~$-\frac{\alpha_\mathrm{S}}{16} \, \cos\theta \left( 2+3\Gamma \left(4-3\cos^2\theta\right) \right) $~~~~~~~  \\
				\cline{2-5}
				& \multicolumn{1}{|c|}{Hard wall}
				& ~~~~~~$ -\frac{3\alpha_\mathrm{S}}{4} \, \cos\theta $~~~~~~
				& $ -\frac{3\alpha_\mathrm{S}}{2} \, \sin\theta $
				& $ -\frac{3\alpha_\mathrm{S}}{8} \, \Gamma \cos\theta \left(1+\sin^2\theta\right)$ \\
				\hline
		\end{tabular}
		\\[5pt]
		\begin{tabular}{|cc|c|c|c|}
		 	\cline{2-5}
			\multicolumn{1}{c}{} & \multicolumn{1}{|c|}{~Interface type~~}
			& $h^2 v_y^\mathrm{HI}$~~ & $h^{3} \Omega_x^\mathrm{HI}$ & $h^{3} \Omega_z^\mathrm{HI}$ \\
			\cline{2-5}
			\multicolumn{1}{c}{}\\[-20pt]\hline
			\multirow{3}{*}{\rotatebox[origin=c]{90}{\makecell{~ \\ Rotlet \\ ~}}}
			& \multicolumn{1}{|c|}{Shear}
			& $-\frac{\alpha_\mathrm{R}}{8} \, \cos\theta $
			& $-\frac{3\alpha_\mathrm{R}}{16} \, \cos\theta $
			& $-\frac{\alpha_\mathrm{R}}{8} \, \sin\theta $ \\
			\cline{2-5}
			& \multicolumn{1}{|c|}{Bending}
			& ~~~~$\frac{\alpha_\mathrm{R}}{8} \, \cos\theta $~~~~
			& $-\frac{\alpha_\mathrm{R}}{16} \, \cos\theta \left( 2-3\Gamma \sin^2\theta \right) $
			& $-\frac{3\alpha_\mathrm{R}}{16} \, \Gamma \sin\theta \cos^2\theta $ \\
			\cline{2-5}
			& \multicolumn{1}{|c|}{Hard wall}
			& 0
			& ~~~~~~$ -\frac{\alpha_\mathrm{R}}{16} \, \cos\theta \left( 5-3\Gamma \sin^2\theta \right)$~~~~~~
			& ~~~~~~$-\frac{\alpha_\mathrm{R}}{16} \, \sin\theta \left( 2+3\Gamma \cos^2\theta \right) $~~~~~~ \\
			\hline
		\end{tabular}
		\caption{
		Expressions of the induced translational and rotational swimming velocities resulting from~$(a)$ Stokeslet and~$(b)$ rotlet near an elastic interface in the quasi-steady limit of vanishing frequency, or equivalently for~$t \rightarrow \infty$.
		The swimming velocities near a no-slip hard wall are obtained by linear superposition of the shear- and bending-related contributions in the vanishing-frequency limit.
    \label{table_vel_wall-Stokeslet-Rotlet}
		}
%	\end{table}
\end{center}
\end{table*}

\begin{table}
\begin{minipage}{\textwidth}
	\hspace{-17.1cm}
	\vspace{-2.2cm} {\Large (a)}
\end{minipage}
\begin{minipage}{\textwidth}
	\hspace{-17.1cm}
	\vspace{-20cm} {\Large (b)}
\end{minipage}
\begin{center}
%	\begin{table}
	\def\arraystretch{3}%
		\begin{tabular}{|cc|c|c|c|}
				\cline{2-5}
					\multicolumn{1}{c}{} & \multicolumn{1}{|c|}{Interface type} & $v_x^\mathrm{HI}$ & $  v_z^\mathrm{HI}$ & $ \Omega_y^\mathrm{HI}$ \\
					\cline{2-5}
					\multicolumn{1}{c}{}\\[-30pt]\hline
					\multirow{2}{*}{Force dipole} & \multicolumn{1}{|c|}{Shear}
					& $\tauS^{-2}$
					& $\tauS^{-3}$
					& $\tauS^{-3}$ \\
					\cline{2-5}
					& \multicolumn{1}{|c|}{Bending}
					& $\tauB^{-4/3}$
					& $\tauB^{-1}$
					& $\tauB^{-4/3}$  \\
					\hline\hline
					\multirow{2}{*}{\makecell{Source dipole / \\ Quadrupole}} & \multicolumn{1}{|c|}{Shear}
					& $\tauS^{-3}$
					& $\tauS^{-4}$
					& $\tauS^{-4}$ \\
					\cline{2-5}
					& \multicolumn{1}{|c|}{Bending}
					& $\tauB^{-4/3}$
					& $\tauB^{-1}$
					& $\tauB^{-4/3}$  \\
					\hline \hline
					\multirow{2}{*}{Stokeslet} & \multicolumn{1}{|c|}{Shear}
					& $\tauS^{-1}$
					& $\tauS^{-3}$
					& $\tauS^{-2}$ \\
					\cline{2-5}
					& \multicolumn{1}{|c|}{Bending}
					& $\tauB^{-1}$
					& $\tauB^{-1/3}$
					& $\tauB^{-1}$  \\
					\hline
		\end{tabular}
		\\[10pt]
		\begin{tabular}{|cc|c|c|c|}
				\cline{2-5}
					\multicolumn{1}{c}{} & \multicolumn{1}{|c|}{Interface type} & $v_y^\mathrm{HI}$ & $  \Omega_x^\mathrm{HI}$ & $\Omega_z^\mathrm{HI}$ \\
					\cline{2-5}
					\multicolumn{1}{c}{}\\[-30pt]\hline
					\multirow{2}{*}{~Rotlet dipole~~} & \multicolumn{1}{|c|}{Shear}
					& $\tauS^{-3}$
					& $\tauS^{-4}$
					& $\tauS^{-4}$ \\
					\cline{2-5}
					& \multicolumn{1}{|c|}{Bending}
					& $\tauB^{-4/3}$
					& $\tauB^{-4/3}$
					& $\tauB^{-5/3}$  \\
					\hline\hline
					\multirow{2}{*}{Rotlet} & \multicolumn{1}{|c|}{Shear}
					& $\tauS^{-2}$
					& $\tauS^{-3}$
					& $\tauS^{-3}$ \\
					\cline{2-5}
					& \multicolumn{1}{|c|}{Bending}
					& $\tauB^{-1}$
					& $\tauB^{-1}$
					& $\tauB^{-4/3}$  \\
					\hline
		\end{tabular}
		\caption{
		Expressions of the long-time decay of the swimming velocities due to $(a)$ dipolar, source dipolar, quadrupolar, and $(b)$~rotlet dipolar hydrodynamic interactions with an elastic interface.
		Here, $\tauS = t/T_\mathrm{S}$ and~$\tauB = t/T_\mathrm{B}$ with~$T_\mathrm{S} = 6\eta h/(B\kS)$ and~$T_\mathrm{B} = 8\eta h^3/\kB$ are characteristic time scales associated with shear and bending deformation modes, respectively. 
		}
		\label{tab:decay-time}
%	\end{table}
\end{center}
\end{table}

\subsubsection{Force quadrupole}

The flow fields generated by a fore–aft asymmetry of the propulsion mechanism can be captured in terms of a \emph{force quadrupole} $\vect{v}_\mathrm{Q} (\R) = \alpha_\mathrm{Q} \bG_\mathrm{Q}(\e,\e,\e)$. Such contributions play a pivotal role for flagellated microorganisms, such as bacteria~\cite{Liao:2007} and sperms~\cite{smith09}, where an asymmetry between the length of the forward-pushing cell and the flagella impacts the propulsive force distribution along the agent and thereby the hydrodynamic flows. resulting effects have been found to induce correlated motion between adjacently swimming bacteria~\cite{Liao:2007}.
It is expected that~$\alpha_\mathrm{Q} >0$ for  microswimmers with large cell bodies and short flagella, while~$\alpha_\mathrm{Q} < 0$ holds for long-flagellated microorganisms with small cell bodies~\cite{spagnolie12, mathijssen16jfm}.

Interestingly, the translational velocity ${v_x}_\mathrm{Q}^\mathrm{HI}$ induced by a force quadrupole parallel to an elastic surface displays at intermediate time scales a weakly non-monotonic behavior before reaching the steady state, see Fig.~\ref{fig:swimmingvel}~$(g)$. In particular, the velocity induced by a surface with pure shear resistance displays the opposite effect than the one induced by a surface with bending resistance at long times considering the present set of parameters.
This implies that, e.g., if bending resistance increases the swimming velocity tangent to the interface, then shear resistance decreases it and vice versa. The induced translational velocity perpendicular to the elastic boundary and the rotational velocity quasi monotonically increase in magnitude over time as resulting from adding both shear and bending contributions, see Fig.~\ref{fig:swimmingvel}~$(h)$ -- $(i)$. 
Notably, the bending effect is once again more pronounced than the one associated with shear.
In the steady state, the translational and rotational velocities approach those induced by a rigid wall, as has been observed for the other higher-order singularity solutions presented above.

Depending on the types of interface, the force quadrupole coefficient, and the pitch angle, quadrupolar hydrodynamic interactions in the steady limit may lead to attraction or repulsion of swimming microorganisms in a complex way.
Considering an interface with only energetic resistance toward shear, we find that ${v_z}_\mathrm{Q}^\mathrm{HI} \propto \alpha_\mathrm{Q} \sin\theta$.
Thus, the swimmer tends to be repelled from the interface when~$\alpha_\mathrm{Q}$ and~$\theta$ have both the same sign, and tends to be attracted toward the interface otherwise.
An analogous discussion holds as well for an interface with only energetic resistance toward bending, or for an interface with both shear and bending deformation modes, provided that $|\theta|< \arccos \left(2\sqrt{15}/15\right)$ in the former, and $|\theta| < \arccos \left(\sqrt{2}/3\right)$ in the latter case.

Next, considering an interface with energetic resistance only toward shear, the rotation rate in the steady state~${\Omega_y}_\mathrm{Q}^\mathrm{HI} \propto \alpha_\mathrm{Q} \cos\theta$.
Thus, the swimmer in the absence of external trapping tends to rotate toward the interface when~$\alpha_\mathrm{Q} > 0$, and away from the interface when~$\alpha_\mathrm{Q} < 0$.
For an elastic interface possessing pure bending resistance, the swimmer may also assume in the steady state an oblique alignment along a pitch angle~$\theta = \pm \theta_\Gamma$, where
\begin{equation}
		\theta_\Gamma =  \arccos \left( \frac{1}{3} \sqrt{8+\frac{11}{\Gamma} - \sqrt{136+\frac{32}{\Gamma}+\frac{121}{\Gamma^2}}} \right) \, . \notag
\end{equation}
Consequently, for~$\alpha_\mathrm{Q}>0$, force quadrupolar hydrodynamic interactions tend to orient the swimmer along~$\theta=-\theta_\Gamma$ when~$\theta<\theta_\Gamma$, and along~$\theta = \pi/2$ otherwise.
In contrast to that, for~$\alpha_\mathrm{Q}<0$, the swimmer tends to be reoriented toward~$\theta=\theta_\Gamma$ when~$\theta>-\theta_\Gamma$, and along~$\theta=-\pi/2$ otherwise. 
An analogous discussion holds when the interface is endowed with both shear and bending resistances in the steady limit (hard wall), where the oblique alignment in this situation is found to be along
\begin{equation}
	\theta_\Gamma = \arccos \left( \sqrt{1+\frac{2}{\Gamma} - \sqrt{\frac{11}{3}+\frac{4}{3\Gamma} + \frac{4}{\Gamma^2}}} \right) \, . \notag \\
\end{equation}

\subsubsection{Rotlet dipole}

In addition, the flow field produced by flagellated microorganisms can be altered by rotation of their body parts, such as the rotation of their flagella bundle and the counter rotation of the cell body in \textit{E. coli} bacteria~\cite{lauga06}. The induced flow far-field can be included at lowest order in terms of a \emph{rotlet dipole}, $\vect{v}_\mathrm{RD} (\R) = \alpha_\mathrm{RD} \, \bG_\mathrm{RD} (\e, \e)$.
A tilted rotlet dipole can conveniently be expanded as a combination of rotlet dipoles orientated parallel and perpendicular to the interface as
\begin{equation}
\begin{split}
 \bG_\mathrm{RD} (\e,\e) &= \bG_\mathrm{RD}(\e_x,\e_x) \cos^2\theta + \bG_\mathrm{RD} (\e_z,\e_z) \sin^2\theta \\
 &\quad+ \bG_\mathrm{RR} (\e_x,\e_z) \sin (2\theta) \, ,
\end{split}
\end{equation}
where $\bG_\mathrm{RR}(\vect{a},\vect{b}) = \big( \bG_\mathrm{RD} (\vect{a}, \vect{b}) + \bG_\mathrm{RD} (\vect{b}, \vect{a})\big)/2$ denotes the symmetric part of the rotlet dipole. Similar to the force quadrupole contribution, the induced swimming velocity parallel to the elastic surface displays a non-monotonic behavior before approaching zero at long times, see Fig.~\ref{fig:swimmingvel}~$(j)$. 
In addition, the shear- and bending-related parts may have opposite contributions to the overall translational velocity tangent to the interface.
At long times, again the velocities of a microswimmer induced by a rigid, no-slip wall are recovered.

Interestingly, the rotation rate around the swimmer body is found to be shear dominated where bending does not play a significant role [Fig.~\ref{fig:swimmingvel}~$(k)$].
Moreover, the rotlet-dipolar hydrodynamic interactions induce a non-vanishing rotation rate about an axis perpendicular to the interface, see Fig.~\ref{fig:swimmingvel}~$(l)$.
This naturally leads in the absence of external trapping to an overall \enquote{swimming in circles}, as has been previously reported for \textit{E. coli} near walls \cite{diluzio05, lauga06} and explained via corresponding theoretical studies that include phenomenological representations of the rotating flagella \cite{dunstan12, daddi18}. As this component vanishes for the other singularities discussed above, we thus expect the introduction of a rotlet dipole to be the simplest possible hydrodynamic modeling of this circling behavior near surfaces.
Remarkably, this rotation rate is independent of the shape factor~$\Gamma$ in the shear-related part but vanishes for a sphere~$(\Gamma=0)$ in the bending-related part.
Considering a swimmer that is aligned parallel to the interface~$(\theta=0)$ in the steady limit, we obtain
\begin{subequations}
	\begin{align}
		\left. {\Omega_z}_\mathrm{RD}^\mathrm{HI} \right|_{\mathrm{S}} &=
		-\frac{3\alpha_\mathrm{RD}}{32h^4} \, , \label{omegaS} \\
		\left. {\Omega_z}_\mathrm{RD}^\mathrm{HI} \right|_{\mathrm{B}} &=
			\frac{3\alpha_\mathrm{RD}}{32h^4} \, \Gamma \, , \label{omegaB} \\
		\left. {\Omega_z}_\mathrm{RD}^\mathrm{HI} \right|_{\mathrm{S} + \mathrm{B}} &=
			-\frac{3\alpha_\mathrm{RD}}{32h^4} \left( 1-\Gamma \right) \, . \label{omegaSundB}
	\end{align}
\end{subequations}
Therefore, assuming that~$\alpha_\mathrm{RD} > 0$, circular motion is expected to be clockwise (when viewed from top) near an interface with pure shear or with both shear and bending rigidities [Eqs.~\eqref{omegaS} and~\eqref{omegaSundB}], and counterclockwise near an interface with pure bending [Eq.~\eqref{omegaB}].
This is in agreement with the behavior observed for a torque-free doublet of counterrotating spheres around its center near an elastic interface~\cite{daddi18coupling}.
It is worth mentioning that, in the steady limit, the system behavior near an interface with pure bending resistance is analogous to that near a flat fluid-fluid interface separating two immiscible fluids with the same viscosity contrast.

%%%%% External forces -  stokeslet & rotlet

\subsection{Contributions due to external forces and torques}

Nature offers a plethora of external stimuli and forces that impact the swimming motion of active agents. Examples include gravitational fields~\cite{campbell13, tenHagen14, campbell17, kuhr17}. % which evoke a gravitactic behavior in certain species including the algae \textit{Chlamydomonas reinhardtii}~\cite{Roberts:2006} and \textit{Paramecium}~\cite{Roberts:2010}, a unicellular type of protozoa. 
The far-field hydrodynamics of externally-trapped self-propelled particles near elastic boundaries can readily be captured in terms of a Stokeslet and rotlet solution to the Stokes equation. The corresponding translational and rotational velocities as functions of time as well as the steady limits are presented in Tabs.~\ref{tab:time-dependent-expressions} and~\ref{table_vel_wall-Stokeslet-Rotlet}.

\subsubsection{Stokeslet}

In the presence of an external force, the \emph{Stokeslet} singularity can be used to capture the associated hydrodynamic flow~\cite{drescher10} and calculate the induced velocity of the microswimmer as $\vect{v}_\mathrm{S} (\R) = \alpha_\mathrm{S} \, \bG (\e)$. Similar as before, a tilted Stokeslet can be decomposed into a superposition of Stokeslets directed parallel and perpendicular to the interface as $\bG(\e ) = \bG (\e_x) \cos\theta + \bG(\e_y) \sin\theta$.
In contrast to the higher-order singularities used to model force-free swimming, the Stokeslet introduces a far field of the fluid flow that decays as $1/h$ and thus represents the leading-order contribution.

In Fig.~\ref{fig:swimmingvel_stokeslet_rotlet}~$(a)$ -- $(c)$, we present the variations of the induced swimming velocities due to a Stokeslet singularity acting near a planar elastic interface with pure shear (green), pure bending (red), or both shear and bending deformation modes (black), using the same parameters as in Fig.~\ref{fig:swimmingvel}.
While resistance toward shear manifests itself in a more pronounced way for the translational motion parallel to the interface, the effect of bending is dominant for the translational motion normal to the interface and for the rotation rate.

In the remainder of our discussion, we assume that the Stokeslet coefficient~$\alpha_\mathrm{S} > 0$.
Correspondingly, the swimmer in the steady state tends to be attracted to the interface when~$\theta>0$, and repelled from it when~$\theta<0$.
Near an interface with resistance only to shear such that~$\Gamma \le 2/3$ (or~$\gamma \le \sqrt{5}$), it follows that ${\Omega_y}_\mathrm{S}^\mathrm{HI} \propto\cos\theta$.
Therefore, the swimmer tends to be reoriented toward the interface~$(\theta=-\pi/2)$.
In contrast to that, for~$\Gamma > 2/3$, the swimmer tends to align along the oblique direction given by~$\theta_\Gamma = \arccos \left( \sqrt{6\Gamma}/(3\Gamma) \right)$ when~$\theta > -\theta_\Gamma$, and along $\theta = -\pi/2$ otherwise.
Near an interface of either pure bending resistance or both shear and bending resistance, ${\Omega_y}_\mathrm{S}^\mathrm{HI} \propto -\cos\theta$, leading to swimmer reorientation away from the interface~$(\theta=\pi/2)$.
Notably, ${\Omega_y}_\mathrm{S}^\mathrm{HI}$ vanishes in the hard-wall limit for a spherical microswimmer~$(\Gamma=0)$.

\subsubsection{Rotlet}

The far-field of an external torque applied to the microswimmer can be described in terms of a \emph{rotlet singularity}.
The rotlet-related contribution to the induced translational velocity resulting from hydrodynamic interactions with the elastic interface has a single non-vanishing component along the~$y$~direction, for which both shear and bending have equal but opposite contributions to the overall dynamics, see Fig.~\ref{fig:swimmingvel_stokeslet_rotlet}~$(d)$.
For the induced rotation rates [Fig.~\ref{fig:swimmingvel_stokeslet_rotlet}~$(e)$ -- $(f)$], the relative importance of shear and bending elasticity depends strongly on the swimmer geometry and orientation.
Analogously to a rotlet dipole, the induced rotational velocity normal to the interface is independent of the shape factor~$\Gamma$ near an interface of only shear resistance, and vanishes for a spherical microswimmer~$(\Gamma = 0)$ near an interface of resistance only to bending.

\subsection{Long-time decay of swimming velocities}

Finally, we briefly comment on the leading-order behavior of the hydrodynamically-induced swimming velocities at long times in approaching the steady limits.
Results are summarized in Tab.~\ref{tab:decay-time} for various singularity and interface types.
For higher-order singularities, the rotation rates are found to decay similarly or much faster than the translational swimming velocities.
Most importantly, the shear-related contributions to the swimming velocities experiences a faster decay in time compared to those related to bending.
Therefore, the system behavior is shear-dominated at early times, while bending is expected to play the more dominant role at later times.

\section{Conclusion}
\label{sec:conclusion}

We have derived exact solutions for the translational and angular velocities of a trapped microswimmer in the vicinity of a deformable surface that features resistance towards bending and shear. Based on far-field calculations we show that the velocities can be decomposed into bending and shear related contributions, which can display opposed behavior, i.e., while one of them enhances the velocities, the other decreases them and vice versa. In particular, the elastic properties of the interface introduce history to the hydrodynamic couplings, which manifests itself in time-dependent translational and rotational velocities of the approaching microswimmer. These velocities strongly depend on the swimming direction, the distance from the interface, the body shape, and details of the swimming mechanism encoded in the singularity coefficients. By accounting for both, bending and shear resistances, the steady state velocities agree with those of an active agent close to a planar, rigid wall.

Our results provide a detailed analysis of far-field hydrodynamic interactions of trapped, self-propelled particles with a deformable surface and are expected to contribute to our understanding of microswimmer motion in their natural surroundings. Based on the proposed theoretical framework, future investigations could elucidate the spatiotemporal behavior of freely moving microswimmers nearby an elastic interface and analyze more closely the potential accumulation of microswimmers at the deformable surface in comparison to a rigid wall~\cite{berke08}. Moreover, an additional, intrinsic curvature of the surface can be included in our model~\cite{daddi17b, spagnolie15}, which could provide a fundamental ingredient for our understanding of microswimmer entrapment and accumulation in realistic biological set-ups.

% \vspace{1.25cm}

\begin{acknowledgments}
	We thank Arnold J.\ T.\ M.\ Mathijssen and Maciej Lisicki for invaluable discussions.
	A.D.M.I, A.M.M., and H.L. gratefully acknowledge support from the DFG (Deutsche Forschungsgemeinschaft) through the projects DA~2107/1-1, ME~3571/2-2, and LO~418/16-3.
	C.K. gratefully acknowledges support from the Austrian Science Fund (FWF) via the Erwin Schr\"{o}dinger Fellowship (Grant No. J~4321-N27).
	A.Z. acknowledges support from the FWF through a Lise Meitner Fellowship (Grant No. M 2458-N36).
	M.M and M.R.A. acknowledge the support of the National Science Foundation (NSF) via Grant No.\@ CMMI-1562871.
	S.G. thanks the Volkswagen Foundation and the DFG  (SFB-TRR 225, subproject B07, 326998133) for financial support."
\end{acknowledgments}

\appendix

\section{\review{Green's functions for a Stokeslet near an elastic interface}}
\label{appendix:greenfunction}

The components of the Green's functions can be expressed in terms of convergent improper (infinite) integrals over the wavenumber and assume the following form 
\begin{align}
\mathcal{G}_{xx}  &= \frac{1}{4\pi}  \int_0^{\infty}  \Intd q \, q \,
\bigg( \tilde{\mathcal{G}}_{+}  J_0 (q \rho_0)
  +   \tilde{\mathcal{G}}_{-} J_2 (q \rho_0) \cos (2\varphi) \bigg) \, ,  \notag  \\
\mathcal{G}_{yy}  &= \frac{1}{4\pi}  \int_0^{\infty} \Intd q \, q \, 
\bigg( \tilde{\mathcal{G}}_{+} J_0 (q \rho_0)
  -   \tilde{\mathcal{G}}_{-} J_2 (q \rho_0 ) \cos (2\varphi) \bigg)  \, , \notag   \\
\mathcal{G}_{zz}  &= \frac{1}{2\pi}
\int_{0}^{\infty}  \Intd q \, q \, 
\tilde{\mathcal{G}}_{zz} J_0 (q \rho_0)  \, , \notag  \\
\mathcal{G}_{xy}  &= \frac{1}{4\pi} \int_0^\infty \Intd q  \, q \, 
\tilde{\mathcal{G}}_{-}  J_2 (q \rho_0) \sin (2\varphi) \, , \notag \\
\mathcal{G}_{rz}  &= \frac{i}{2\pi} \int_{0}^{\infty} \Intd q \, q \, 
\tilde{\mathcal{G}}_{lz} J_1 (q \rho_0)  \, , \notag \\
\mathcal{G}_{zr}  &= \frac{i}{2\pi} \int_{0}^{\infty} \Intd q \, q \, 
\tilde{\mathcal{G}}_{zl} J_1 (q \rho_0)  \, , \notag
\end{align}
wherein $\rho_0 = \sqrt{(x-x_0)^2 + (y-y_0)^2}$ denotes the radial distance and $\varphi := \arctan ((y-y_0)/(x-x_0))$ is the azimuthal angle (c.f.\ inset of Fig.~\ref{illustration-of-system-setup}).
Here $J_n(\cdot)$ represents the $n$-th order Bessel function of the first kind~\cite{abramowitz72} and we introduce
\begin{equation}
\tilde{\mathcal{G}}_{\pm} (q,z,\omega) := \tilde{\mathcal{G}}_{tt}(q,z,\omega) \pm \tilde{\mathcal{G}}_{ll}(q,z,\omega) \, , \notag
\end{equation}
with
\begin{align}
\tilde{\mathcal{G}}_{ll}  &= \frac{1}{4 \eta q}
\bigg(
(1-q |z - h|) e^{-q|z-h|}   \notag \\
&\quad+ \left( \frac{2i qh (1-q h)(1-qz)}{\beta - 2i qh} + \frac{8i q^5 z h^4}{\betaB^3 - 8i (qh)^3} \right) e^{-q(z+h)}
\bigg) \, ,
\notag
\\
\tilde{\mathcal{G}}_{tt}  &= \frac{1}{2 \eta q} \left( e^{-q|z-h|} + \frac{i B qh}{\beta - i B qh} e^{-q(z+h)}  \right) \, .
\notag
\end{align}
The remaining Green's functions in Fourier space read
\begin{align}
\tilde{\mathcal{G}}_{zz}  &= \frac{1}{4 \eta q}
\bigg(
\left( 1+q|z - h| \right) e^{-q|z-h|}   \notag \\
&\quad+ \left( \frac{2i q^3 zh^2}{\beta - 2iqh} + \frac{8i (qh)^3 (1+qz)(1+q h)}{\betaB^3-8i (qh)^3} \right) e^{-q(z+h)}
\bigg) \, ,
\notag  \\
\tilde{\mathcal{G}}_{lz}   &= \frac{i}{4 \eta q}
\bigg(
-q (z - h) e^{-q|z-h|}  \notag \\
&\quad+ \bigg( \frac{2i (qh)^2 (1-qz)}{\beta-2i qh}
- \frac{8i q^4 zh^3 (1+q h)}{\betaB^3 - 8i (qh)^3} \bigg) e^{-q(z+h)}
\bigg) \, , \nonumber \\
\tilde{\mathcal{G}}_{zl}  &= \frac{{i}}{4 \eta q}
\bigg(
-q(z - h) e^{-q|z-h|}  \notag \\
&\quad+ \bigg(- \frac{2i q^2zh (1-q h)}{\beta - 2i qh}
+ \frac{8i q^4 h^4 (1+qz)}{\betaB^3 - 8i (qh)^3} \bigg) e^{-q(z+h)}
\bigg) \, . \nonumber
\end{align}

The Green's functions comprise both bulk contributions and the frequency-dependent corrections due to the presence of the elastic interface.
The terms involving $\beta$ and $\betaB$ are, respectively, contributions associated with shear and bending.
Moreover, the remaining components of the Green's functions can readily be obtained from the usual transformation relations.
Specifically, this means $\mathcal{G}_{xz} = \mathcal{G}_{rz}\cos\varphi$, $\mathcal{G}_{yz} = \mathcal{G}_{rz}\sin\varphi$, $\mathcal{G}_{zx} = \mathcal{G}_{zr}\cos\varphi$, $\mathcal{G}_{zy} = \mathcal{G}_{zr}\sin\varphi$, and $\mathcal{G}_{yx} = \mathcal{G}_{xy}$. In the quasi-steady limit of vanishing frequency ($\beta=\betaB=0$), the Green's functions reduce to the well-known Blake tensor near a no-slip wall~\cite{blake71, blake74}. Physically, this limit corresponds to an infinitely stiff wall, for which the displacement field at the interface identically vanishes.

\section{\review{Higher-order singularities in an unbounded fluid domain}}
\label{appendix:higher-order-singularities}

\review{In this Appendix, we provide for completeness analytical expressions of the higher-order Stokes singularities in an unbounded fluid domain, i.e., in the absence of the confining elastic interface.
By making use of the analytical recipes introduced in Sec.~\ref{subsec:multipoleExpansion}, we readily obtain}
\begin{align}
		\bG_\mathrm{R}^\infty &=  \frac{1}{s^2} \left( \e \times \vect{\hat{s}} \right) \, , \notag \\
		\bG_\mathrm{D}^\infty &=  \frac{1}{s^2}
			\left( 3\left( \e \cdot \vect{\hat{s}} \right)^2-1 \right) \vect{\hat{s}} \, , \notag  \\
		\bG_\mathrm{SD}^\infty &= \frac{1}{s^3}
		\left( 3 \left( \e \cdot \vect{\hat{s}}  \right) \vect{\hat{s}} - \e\right) \, ,  \notag \\
		\bG_\mathrm{Q}^\infty &= \frac{1}{s^3}
			\bigg( 3\left( 5 \left( \e\cdot \vect{\hat{s}} \right)^3-3\left( \e\cdot \vect{\hat{s}} \right) \right) \vect{\hat{s}}
			- \left( 3\left( \e\cdot \vect{\hat{s}} \right)^2-1 \right) \e	\bigg)  \, , \notag \\
		\bG_\mathrm{RD}^\infty &= \frac{3}{s^3} \left( \e \cdot \vect{\hat{s}}  \right)
		\left( \e \times \vect{\hat{s}}  \right) \, , \notag
\end{align}
\review{where, again, $\vect{s} = \R-\R_0$ denotes the position vector relative to the singularity location, $s=|\vect{s}|$, $\hat{\vect{s}} = \vect{s}/s$, and~$\e$ stands for the orientation unit vector of the swimmer as defined by Eq.~\eqref{Orientierung} of the main body of the paper.
Notably, the rotlet (R) and force dipole (D) decay in the far-field limit as~$1/s^2$, whereas the source dipole (SD), force quadrupole (Q), and rotlet dipole (RD) undergo a faster decay as~$1/s^3$.}

\section{\review{Expression of the induced-swimming velocities in the frequency and temporal domains}}\label{appendix:tables}

% To change the numbering style of tables in Appendix
\setcounter{table}{0}
\renewcommand{\thetable}{C\arabic{table}}

\review{
Here, we present the main mathematical expressions obtained in this paper in the form of tables. 
We provide in tables~\ref{tab:freq-dependent-expressions} and~\ref{tab:coeff-freq} explicit analytical expressions of the frequency-dependent translational swimming velocities and rotation rates resulting from the fluid-mediated hydrodynamic interactions with a nearby planar elastic interface.
In tables~\ref{tab:time-dependent-expressions} and \ref{tab:coeff}, we list the corresponding expressions in the temporal domain for the start-up motion from static conditions.
As already mentioned in the main text, only the induced translational swimming velocities in the temporal domain are provided.
The rotation rates have rather lengthy and complex analytical expressions and thus are not listed here.}

\begin{table*}
	\def\arraystretch{3}%
	\review{
	\begin{tabular}{|c|c|c|}
			\hline
			\multirow{3}{*}{\rotatebox[origin=c]{90}{\makecell{~ \\ Force dipole \\ ~}}} & ${v_x}_\mathrm{D}^\mathrm{HI}$ & \(\displaystyle \frac{\alpha_\mathrm{D} \sin (2\theta)}{2h^2}  \int_0^\infty \mkern-12mu \Intd u
			\left( \frac{ N_\mathrm{D}^\mathrm{S}}{S} + \frac{8u^6}{8u^3+i\betaB^3} \right)e^{-2u} \) \\
			& ${v_z}_\mathrm{D}^\mathrm{HI}$ & \(\displaystyle \frac{\alpha_\mathrm{D}\left(2-3\cos^2\theta\right)}{h^2} \int_0^\infty \mkern-12mu \Intd u
			\left( \frac{u^3(u-1)}{2u+i\beta} + \frac{4u^5(u+1)}{8u^3+i\betaB^3} \right) e^{-2u} \) \\
			& ${\Omega_y}_\mathrm{D}^\mathrm{HI}$ & \(\displaystyle \frac{\sin (2\theta)}{24 h^3} \int_0^\infty \mkern-12mu \Intd u
			\bigg( \frac{u^3}{S} \left(H^\mathrm{S}_\mathrm{D} + A^\mathrm{S}_\mathrm{D}\cos^2\theta\right) % \notag \\
			% &\quad
			+ \frac{12u^6}{8u^3 + i\betaB^3} \left( 8+\Gamma u \left(4-3\cos^2\theta\right) \right)  \bigg) e^{-2u} \) \\
			\hline
			\multirow{3}{*}{\rotatebox[origin=c]{90}{\makecell{~ \\ Source dipole \\ ~}}} & ${v_x}_\mathrm{SD}^\mathrm{HI}$ & \(\displaystyle -\frac{\alpha_\mathrm{SD} \cos\theta}{h^3} \int_0^\infty \mkern-12mu \Intd u
			\left( \frac{N^\mathrm{S}_\mathrm{SD}}{S}+\frac{4u^6}{8u^3+i\betaB^3} \right) e^{-2u} \)  \\
			& ${v_z}_\mathrm{SD}^\mathrm{HI}$ & \(\displaystyle -\frac{\alpha_\mathrm{SD}\sin\theta}{h^3} \int_0^\infty \mkern-12mu \Intd u \,
			\left( \frac{2u^4}{2u+i\beta} + \frac{8u^5 (1+u)}{8u^3+i\betaB^3}\right) e^{-2u}  \) \\
			& ${\Omega_y}_\mathrm{SD}^\mathrm{HI}$ & \(\displaystyle -\frac{\alpha_\mathrm{SD}\cos\theta}{h^4}  \int_0^\infty \mkern-12mu \Intd u \,
			\bigg(	\frac{u^4}{S} \left(H^\mathrm{S}_\mathrm{SD} + A^\mathrm{S}_\mathrm{SD} \cos^2\theta\right) % \notag \\
			% &\quad
			+\frac{4 u^6  }{8u^3+i\betaB^3} \left( 1+\Gamma u \left(2-\cos^2\theta \right) \right)
			\bigg)  e^{-2u}  \) \\
			\hline
			\multirow{3}{*}{\rotatebox[origin=c]{90}{\makecell{~ \\ Quadrupole \\ ~}}}  & ${v_x}_\mathrm{Q}^\mathrm{HI}$ & \(\displaystyle \frac{\alpha_\mathrm{Q}\cos\theta}{4 h^3} \int_0^\infty \mkern-12mu \Intd u \,
			\bigg( \frac{ u^3 }{S} \left(N^\mathrm{S}_\mathrm{Q} + M^\mathrm{S}_\mathrm{Q} \cos^2\theta\right) % \notag \\
			% &\quad
			+  \frac{4u^6}{8u^3+i\betaB^3} \left(4u (2-3u) + (15u-8) \cos^2\theta\right)
			\bigg) e^{-2u}  \) \\
			& ${v_z}_\mathrm{Q}^\mathrm{HI}$ & \(\displaystyle \frac{\alpha_\mathrm{Q}\sin\theta}{h^3}  \int_0^\infty \mkern-12mu \Intd u \,
			\bigg( \frac{ u^4 }{2u+i\beta} \left(2 (2-u)+ (5u-8) \cos^2\theta\right) % \notag \\
			% &\quad
			+  \frac{4u^5 (1+u)}{8u^3+i\betaB^3} \left( 2 (1-u) + (5u-3) \cos^2\theta\right)
			\bigg) e^{-2u}   \) \\
			& ${\Omega_y}_{\mathrm{Q}}^\mathrm{HI}$ & \(\displaystyle \frac{\alpha_\mathrm{Q}\cos\theta}{8 h^4}  \int_0^\infty \mkern-12mu \Intd u \,
			\bigg( \frac{ u^4}{S} \left(W^\mathrm{S}_\mathrm{Q} \cos^4\theta + A^\mathrm{S}_\mathrm{Q}  \cos^2\theta+ H^\mathrm{S}_\mathrm{Q}\right) % \notag \\
			% &\quad
			+ \frac{u^6}{8u^3+i\betaB^3} \left( W^\mathrm{B}_\mathrm{Q} \cos^4\theta + A^\mathrm{B}_\mathrm{Q}  \cos^2\theta+ H^\mathrm{B}_\mathrm{Q}\right)
			\bigg) e^{-2u}  \) \\
			\hline
			\multirow{3}{*}{\rotatebox[origin=c]{90}{\makecell{~ \\ Rotlet dipole \\ ~}}} & ${v_y}_\mathrm{RD}^\mathrm{HI}$ & \(\displaystyle \frac{\alpha_\mathrm{RD} \sin (2\theta)}{h^3} \int_0^\infty  \mkern-12mu \Intd u \,
			\bigg(
			\frac{N^\mathrm{S}_\mathrm{RD}}{4S} % \\
			-\frac{2u^6}{8u^3+i\betaB^3} \bigg) e^{-2u}  \) \\
			& ${\Omega_{x}}_\mathrm{RD}^\mathrm{HI}$ & \(\displaystyle \frac{\alpha_\mathrm{RD} \sin (2\theta)}{16 h^4}  \int_0^\infty  \mkern-12mu \Intd u \,
			\bigg(
			\frac{u^4}{S} \left(G^\mathrm{S}_\mathrm{RD} + K^\mathrm{S}_\mathrm{RD} \cos^2\theta\right) % \notag \\
			% &\quad
			+  \frac{8u^6}{8u^3+i\betaB^3} \left(4 (1-\Gamma u) + 3\Gamma u \cos^2\theta\right)
			\bigg) e^{-2u} \) \\
			& ${\Omega_{z}}_\mathrm{RD}^\mathrm{HI}$ & \(\displaystyle \frac{\alpha_\mathrm{RD}}{8 h^4} \int_0^\infty \mkern-12mu \Intd u \,
			\bigg( \frac{ u^4}{S} \left(W^\mathrm{S}_\mathrm{RD} \cos^4\theta+ A^\mathrm{S}_\mathrm{RD}\cos^2\theta + H^\mathrm{S}_\mathrm{RD}\right) % \notag \\
			% &\quad
			+\frac{8\Gamma u^7 }{8u^3+i\betaB^3} \left( \left( 4 - 3 \cos^2\theta \right) \cos^2\theta\right)
			\bigg) e^{-2u} \) \\
			\hline\hline
			\multirow{3}{*}{\rotatebox[origin=c]{90}{\makecell{~ \\ Stokeslet \\ ~}}} & ${v_x}_\mathrm{S}^\mathrm{HI}$ & \(\displaystyle
	 		-\frac{\alpha_\mathrm{S} \cos\theta}{h}
			\int_0^\infty \Intd u \left(
			\frac{N_\mathrm{S}^\mathrm{S}}{S} + \frac{4u^5}{8u^3+i\betaB^3}	\right) e^{-2u} \) \\
			& ${v_z}_\mathrm{S}^\mathrm{HI}$ & \(\displaystyle
			-\frac{\alpha_\mathrm{S} \sin\theta}{h}
			\int_0^\infty \Intd u \left(
			\frac{2u^3}{2u+i\beta} + \frac{8u^3 \left(u+1\right)^2}{8u^3+i\betaB^3}
			\right) e^{-2u} \) \\
			& ${\Omega_y}_\mathrm{S}^\mathrm{HI}$ & \(\displaystyle
			-\frac{\alpha_\mathrm{S} \cos\theta}{h^2} \int_0^\infty \Intd u  \,
			\bigg( \frac{u^2}{2S} \left( H_\mathrm{S}^\mathrm{S} + A_\mathrm{S}^\mathrm{S} \cos^2\theta \right) % \notag \\
			% &\quad
			+ \frac{4u^5}{8u^3+i\betaB^3} \left( 1+2\Gamma u + 3\Gamma - \Gamma (u+3)\cos^2\theta \right) \bigg) e^{-2u} \) \\
			\hline
			\multirow{3}{*}{\rotatebox[origin=c]{90}{\makecell{~ \\ Rotlet \\ ~}}}& ${v_y}_\mathrm{R}^\mathrm{HI}$ & \(\displaystyle
			\frac{\alpha_\mathrm{R} \cos\theta}{2h^2}
			\int_0^\infty \Intd u
			\left( \frac{H_\mathrm{R}^\mathrm{S}}{S} + \frac{8u^5}{8u^3+i\betaB^3} \right)e^{-2u} \) \\
			& ${\Omega_x}_\mathrm{R}^\mathrm{HI}$ & \(\displaystyle
			\frac{\alpha_\mathrm{R} \cos\theta}{h^3} \int_0^\infty \Intd u \, \bigg(
			\frac{u^3}{4S} \left( G_\mathrm{R}^\mathrm{S} + K_\mathrm{R}^\mathrm{S} \cos^2\theta \right) % \notag \\
			% &\quad
			-\frac{4u^5}{8u^3+i\betaB^3} \left( 1-\Gamma u  + \Gamma u \cos^2\theta \right)
			\bigg) e^{-2u} \) \\
			& ${\Omega_z}_\mathrm{R}^\mathrm{HI}$ & \(\displaystyle
			\frac{\alpha_\mathrm{R} \sin\theta}{h^3} \int_0^\infty \Intd u \, \bigg(
			\frac{u^3}{4S} \left( H_\mathrm{R}^\mathrm{S} + K_\mathrm{R}^\mathrm{S} \cos^2\theta \right) % \notag \\
			% &\quad
			-\frac{4u^6}{8u^3+i\betaB^3} \, \Gamma  \cos^2\theta \bigg)
			e^{-2u} \) \\
			\hline
	\end{tabular}
	}
	\caption{\review{Expressions of the frequency-dependent evolutions of the induced-swimming velocities resulting from hydrodynamic interactions with the elastic interface. 
	Here, we have used the abbreviation $S = 2Bu^2 + (B+2)i\beta u - \beta^2$.}}
					\label{tab:freq-dependent-expressions}
\end{table*}

\begin{table*}
	\def\arraystretch{2}%
	\review{
	\begin{tabular}{|c|c|c|}
	\hline
		% \multirow{3}{*}{\rotatebox[origin=c]{90}{\makecell{~ \\ Force dipole \\ ~}}}
		\multirow{3}{*}{\rotatebox[origin=c]{0}{\makecell{~D~~}}} &$N_\mathrm{D}^\mathrm{S}$ & \(\displaystyle
		  u^2 \left(\left( 2u^2-4u+B+2 \right) i\beta + 2Bu\left( u^2-2u+2 \right) \right) \) \\
		  &$H^\mathrm{S}_\mathrm{D}$ & \(\displaystyle
		  3\Gamma \left( 2\left( 2u^2-4u+2-B \right)i\beta + 4Bu^2 (u-2) \right)   %\notag \\
		  % &\quad
		  + 6 \left( (4u-B-4)i\beta + 2Bu (2u-3) \right) \) \\
		  &$A^\mathrm{S}_\mathrm{D}$ & \(\displaystyle
		  3\Gamma \left( 3 (-u^2+2u+B-1)i\beta + 3Bu (-u^2+2u+1) \right)   \) \\
		 \hline
		 % \multirow{3}{*}{\rotatebox[origin=c]{90}{\makecell{~ \\ Source dipole \\ ~}}}
		 \multirow{3}{*}{\rotatebox[origin=c]{0}{\makecell{SD}}}
		 &$N^\mathrm{S}_\mathrm{SD}$ & \(\displaystyle u^3(Bu+i\beta)(u-1) \) \\
		 &$H^\mathrm{S}_\mathrm{SD}$ & \(\displaystyle 2 \left(Bu+i\beta\right)\left(1+2\Gamma (u-1)\right) \) \\
		 &$A^\mathrm{S}_\mathrm{SD}$ & \(\displaystyle - \left(Bu+i\beta\right) \Gamma (u-1) \) \\
		 \hline
		 %\multirow{8}{*}{\rotatebox[origin=c]{90}{\makecell{~ \\ Quadrupole \\ ~}}}
		 \multirow{8}{*}{\rotatebox[origin=c]{0}{\makecell{~Q~~}}}
		 &$N^\mathrm{S}_\mathrm{Q}$ & \(\displaystyle -\big( 4i\beta \big( 3u^2 -8u+5+B \big) % \notag  \\
	   + 4Bu \big( 3u^2-8u+7 \big) \big) \) \\
	   &$M^\mathrm{S}_\mathrm{Q}$ & \(\displaystyle \left( 15u^2-38u+5B+23 \right)i\beta % \notag \\
	   + Bu \big( 15u^2- 38u+33 \big) \) \\
		&$W^\mathrm{S}_\mathrm{Q}$ & \(\displaystyle \Gamma \left( 6i\beta(B-2-u^2+3u)-6Bu^2(u-3) \right) \) \\
		&$A^\mathrm{S}_\mathrm{Q}$ & \(\displaystyle ~~~2Bu(15u-28) + \Gamma  \big( i\beta(26-9B+12u^2-38u) +2Bu(6u^2-19u+4) \big) + i\beta (30u-5B-46)~~~ \)  \\
		 	&$H^\mathrm{S}_\mathrm{Q}$ & \(\displaystyle \Gamma \left( 4i\beta(2u+B-2)+8Bu^2 \right) -24Bu(u-2) % \notag \\
		 	% &\quad
		 	+ 4i\beta (10+B-6u) \) \\
		 	&$W^\mathrm{B}_\mathrm{Q}$ & \(\displaystyle 24\Gamma u(1-u) \)  \\
		 	&$A^\mathrm{B}_\mathrm{Q}$ & \(\displaystyle 8 \left( 15u-8+\Gamma u(6u-7) \right) \)  \\
		 	&$H^\mathrm{B}_\mathrm{Q}$ & \(\displaystyle 32 (2-3u+\Gamma u) \) \\
		 \hline
		 % \multirow{6}{*}{\rotatebox[origin=c]{90}{\makecell{~ \\ Rotlet dipole \\ ~}}}
		 \multirow{6}{*}{\rotatebox[origin=c]{0}{\makecell{RD}}}
		 &$N^\mathrm{S}_\mathrm{RD}$ & \(\displaystyle 2u^3 \left( (1+B-u)i\beta + Bu(3-u) \right) \) \\
		 	&$K^\mathrm{S}_\mathrm{RD}$ & \(\displaystyle \Gamma \left(6Bu(u-2)-3i\beta(2+B-2u)\right) \) \\
		 	&$G^\mathrm{S}_\mathrm{RD}$ & \(\displaystyle \Gamma \left( 4i\beta(2+B-2u)-8Bu(u-2)\right)  % \notag \\
		 	% &\quad
		 	+16Bu + 4i\beta (2+B) \) \\
		 	&$W^\mathrm{S}_\mathrm{RD}$ & \(\displaystyle \Gamma \left( 3i\beta(2+B-2u)+6Bu(2-u) \right) \) \\
		 	&$A^\mathrm{S}_\mathrm{RD}$ & \(\displaystyle \Gamma \left( 4i\beta(2u-2-B)+8Bu(u-2) \right)  % \notag \\
		 	% &\quad
		 	-6B(2u+i\beta) \)  \\
		 	&$H^\mathrm{S}_\mathrm{RD}$ & \(\displaystyle 4B (2u+i\beta) \) \\
		\hline
		\hline
		%\multirow{3}{*}{\rotatebox[origin=c]{90}{\makecell{~ \\ Stokeslet \\ ~}}}
		\multirow{3}{*}{\rotatebox[origin=c]{0}{\makecell{S}}}
		 &$N_\mathrm{S}^\mathrm{S}$ & \(\displaystyle
			u \left( \left(u^2-2u+B+1\right)i\beta + Bu \left(u^2-2u+3\right) \right) \) \\
		&$H_\mathrm{S}^\mathrm{S}$ & \(\displaystyle
		\Gamma \left( \left( 4u^2-2u-B-2 \right)i\beta+2Bu \left(2u^2-u-2\right) \right) % \notag \\
				% &\quad
				+ \left(2u-2-B\right)i\beta + 2Bu \left(u-2\right) \) \\
		&$A_\mathrm{S}^\mathrm{S}$ & \(\displaystyle
		-\Gamma \left( 2\left(u^2+u-2-B\right)i\beta + 2Bu \left(u^2+u-4\right) \right) \) \\
		\hline
		% \multirow{4}{*}{\rotatebox[origin=c]{90}{\makecell{~ \\ Rotlet \\ ~}}}
		\multirow{4}{*}{\rotatebox[origin=c]{0}{\makecell{R}}}
		&$H_\mathrm{R}^\mathrm{S}$ & \(\displaystyle
		u^2 \left( \left(2u-2-B\right)i\beta+2Bu\left(u-2\right) \right) \) \\
		&$K_\mathrm{R}^\mathrm{S}$ & \(\displaystyle
		\Gamma \left( \left(-4u+B+4\right)i\beta - 2Bu \left(2u-3\right) \right)  \) \\
		&$G_\mathrm{R}^\mathrm{S}$ & \(\displaystyle
		-K_\mathrm{R}^\mathrm{S} -6Bu-(B+4)i\beta  \) \\
		&$H_\mathrm{R}^\mathrm{S}$ & \(\displaystyle
		-2B\left(2u+i\beta\right)  \) \\
		\hline
	\end{tabular}
	}
	\caption{\review{Expressions of the frequency-dependent coefficients appearing in~Tab.~\ref{tab:freq-dependent-expressions}.}
			}
			\label{tab:coeff-freq}
\end{table*}

\begin{table*}\raggedleft % to leave space for the labels (a) und (b)
\iffalse
			\begin{minipage}{\textwidth}
				\hspace{-18.5cm}
				\vspace{-1.4cm} {\Large (a)}
			\end{minipage}
			\begin{minipage}{\textwidth}
				\hspace{-18.5cm}
				\vspace{-27.5cm} {\Large (b)}
			\end{minipage}
\fi
\begin{center}
	\def\arraystretch{2.75}
	\begin{tabular}{|c|c|c|}
			\hline
			\multirow{2}{*}{\rotatebox[origin=c]{90}{\makecell{~ \\ Force dipole \\ ~}}}
			& ${v_x}_\mathrm{D}^\mathrm{HI} $ & \(\displaystyle \frac{\alpha_\mathrm{D} \sin(2\theta)}{2h^2} \left(
				\frac{\tauS \, J_\mathrm{D}^\mathrm{S}}{8\left(1+\tauS\right)^4 \left(2+B\tauS\right)^2}
				+  \int_0^\infty \mkern-12mu \Intd u \,  u^3 \xi(u) \right) \)
				\\
			& ${v_z}_\mathrm{D}^\mathrm{HI} $ & \(\displaystyle
			-\frac{\alpha_\mathrm{D} \left(2-3\cos^2\theta\right)}{2h^2} \left( \frac{\tauS \left( \tauS^3+4\tauS^2+6\tauS+6 \right)}{8 \left( \tauS + 1 \right)^4}
			+ \int_0^\infty \mkern-12mu \Intd u \, (1+u)u^2 \xi(u) \right)
			\) \\
			\hline
			\multirow{2}{*}{\rotatebox[origin=c]{90}{\makecell{~ \\ Source dipole \\ ~}}}
			& ${v_x}_\mathrm{SD}^\mathrm{HI}$  & \(\displaystyle-\frac{\alpha_\mathrm{SD}\cos\theta}{2h^3} \left(
				\frac{\tauS \left( \tauS^3+4\tauS^2+6\tauS+6 \right)}{8 \left(1+\tauS\right)^4}
				+
					\int_0^\infty \mkern-12mu \Intd u \, u^3 \xi(u) \right)
				\)\\
			& ${v_z}_\mathrm{SD}^\mathrm{HI}$  & \(\displaystyle
			-\frac{\alpha_\mathrm{SD}\sin\theta}{h^3}  \left(
				\frac{3\tauS \left(2+\tauS\right)\left(\tauS^2+2\tauS+2\right)}{8\left(1+\tauS\right)^4}
				+
					\int_0^\infty \mkern-12mu \Intd u \, (1+u)u^2 \xi(u) \right)
				\) \\
			\hline
			\multirow{2}{*}{\rotatebox[origin=c]{90}{\makecell{~ \\ Quadrupole \\ ~}}}
			& ${v_x}_\mathrm{Q}^\mathrm{HI}$   & ~~~\(\displaystyle \frac{\alpha_\mathrm{Q}\cos\theta}{8h^3} \left(
				\frac{\tauS}{4\left(1+\tauS\right)^5 \left(2+B\tauS\right)^3}
				\left( Y_\mathrm{Q}^\mathrm{S} \cos^2\theta - J_\mathrm{Q}^\mathrm{S} \right)
				+
					\int_0^\infty \mkern-12mu \Intd u \, u^3 \, \left( 2-2u + \left(5u-3\right) \cos^2\theta \right) \xi(u) \right)
				\)~~~ \\
			& ${v_z}_\mathrm{Q}^\mathrm{HI}$  & \(\displaystyle \frac{\alpha_\mathrm{Q}\sin\theta}{2h^3} \left(
				\frac{3\tauS}{4\left(1+\tauS\right)^5}
				\left( R_\mathrm{Q}^\mathrm{S} \cos^2\theta -2 \right)
				+
					\int_0^\infty \mkern-12mu \Intd u \, (1+u) \, u^2 \, \left( 8-12u+\left(15u-8\right)\cos^2\theta \right) \xi(u) \right)
				\) \\
			\hline
			\multirow{1}{*}{\rotatebox[origin=c]{90}{\makecell{~ \\ Rotlet \\ dipole \\ ~}}}
			& ${v_y}_\mathrm{RD}^\mathrm{HI}$  & \(\displaystyle \frac{\alpha_\mathrm{RD} \sin(2\theta)}{4h^3} \left(
				\frac{\tauS J_\mathrm{RD}^\mathrm{S}}{8 \left(1+\tauS\right)^4 \left(2+B\tauS\right)^3}
				-
					\int_0^\infty \mkern-12mu \Intd u \, u^3 \xi(u) \right)
				\) \\
				\hline\hline
				\multirow{2}{*}{\rotatebox[origin=c]{90}{\makecell{~ \\ Stokeslet \\ ~}}}
			& ${v_x}_\mathrm{S}^\mathrm{HI}$ & \(\displaystyle -\frac{\alpha_\mathrm{S}\cos\theta}{2h}
			\left( \frac{\tauS J_\mathrm{S}^\mathrm{S}}{4\left(1+\tauS\right)^3\left(2+B\tauS\right)}
			+ \int_0^\infty \Intd u \, u^2\xi(u) \right)
			\) \\
			& ${v_z}_\mathrm{S}^\mathrm{HI} $ & \(\displaystyle -\frac{\alpha_\mathrm{S}\sin\theta}{h}
			\left( \frac{\tauB \left( \tauB^2+3\tauB+3 \right)}{4 \left(1+\tauB\right)^3}
			+
			 \int_0^\infty \Intd u \,  (1+u)^2 \xi(u) \right)
			\) \\
			\hline
			\multirow{1}{*}{\rotatebox[origin=c]{90}{\makecell{~ \\ Rotlet \\ ~}}}
			& ${v_y}_\mathrm{R}^\mathrm{HI} $ &
			\(\displaystyle
			\frac{\alpha_\mathrm{R} \cos\theta}{h^2} \left( -\frac{\tauS J_\mathrm{R}^\mathrm{S}}{8\left(1+\tauS\right)^3 \left(2+B\tauS\right)^2}
			+\int_0^\infty \Intd u \, u^2 \xi(u) \right)
			 \) \\
			 \hline
		\end{tabular}
		\caption{Expressions of the time-dependent evolutions of the induced-swimming velocities due to hydrodynamic interactions with the elastic interface. 
		Here, $\xi(u) = e^{-2u} - e^{-2u \left( 1+\tauB u^2 \right)}$ is a bending-related dimensionless function.
				}
				\label{tab:time-dependent-expressions}
\end{center}
\end{table*}
% ~\\
% \vspace{0.25cm}
\begin{table*}
\begin{center}
		\def\arraystretch{2}%
		\begin{tabular}{|c|c|}
		\hline
		%\multirow{1}{*}{\rotatebox[origin=c]{0}{\makecell{ ~D~ }}}
		% & 
		$J_\mathrm{D}^\mathrm{S}$ & \(\displaystyle 3B^2 \tauS^5 + 12B(1+B)\tauS^4+4 \left(1+4B(B+3)\right)\tauS^3
		% &\quad
		+ 4 \left(4+B\left(3B+16\right)\right) \tauS^2 + 2\left(8+B\left(B+24\right)\right) \tauS
		% &\quad
		+ 8 \left(B+2\right) \) \\
		\hline
		% \multirow{5}{*}{\rotatebox[origin=c]{0}{\makecell{ ~Q~ }}}
		% & 
		\multirow{2}{*}{$Y_\mathrm{Q}^\mathrm{S}$} & $21B^3\tauS^7 + 21B^2 \left(6+5B\right)\tauS^6
		+ 42B\left(5B+6\left(B+3\right)\right) \tauS^5 % \notag \\
		% &\quad
		+ \left( 187B^3+1260B\left(B+1\right)+88 \right) \tauS^4
		\, + \, 2\big(58B^3+561B^2 $ \\
		% & 
		&$+\, 1260B+220\big) \tauS^3
		+ 2 \left(5B^3+348B^2+1122B+440\right) \tauS^2 % \notag \\
		% &\quad
		+ 12 \left(5B^2+116B+58\right) \tauS
		+ 24 \left(5B+22\right) $ \\
		% & 
		\multirow{2}{*}{$J_\mathrm{Q}^\mathrm{S}$} & $16B^3 \tauS^7 + 16B^2 \left(5B+6\right)\tauS^6
		+32B \left(5B^2+15B+6\right) \tauS^5 % \notag  \\
		% &\quad
		+4 \left(35B^3+240B \left(B+1\right)+16\right) \tauS^4
		+8 \big( 11B^3+105B^2 $ \\
		% & 
		& $+ \,240B+40 \big) \tauS^3
		+8 \left(B^3 + 66B^2 + 210B+80\right) \tauS^2 % \notag \\
		% &\quad
		+48 \left(B^2+22B+10\right) \tauS + 96 \left(B+4\right) $ \\
		% & 
		$R_\mathrm{Q}^\mathrm{S}$ & $\tauS^4+5\tauS^3+10\tauS^2+10\tauS+9 $ \\
		\hline
		% \multirow{2}{*}{\rotatebox[origin=c]{0}{\makecell{ ~RD~ }}}
		 \multirow{2}{*}{$J_\mathrm{RD}^\mathrm{S}$} 
		& $3B^3 \tauS^6 + 6B^2 (2B+3) \tauS^5
		+18B \left(B^2+4B+2\right) \tauS^4 % \notag \\
		% &\quad
		+2 \left( 5B^3+54B^2+72B-4 \right) \tauS^3
		+ 4 \, \big( B^3 + 15B^2 $ \\
		% & 
		& $+\, 54B-8 \big) \tauS^2
		+24 \left(B^2+5B-2\right) \tauS + 48 \left(B-1\right) $ \\
		\hline\hline
		% \multirow{1}{*}{\rotatebox[origin=c]{0}{\makecell{ ~S~ }}}
		$J_\mathrm{S}^\mathrm{S}$ 
		& $5B\tauS^3+\left(2+13B\right)\tauS \left(1+\tauS\right) + 2\left(1+2B\right) $ \\
		\hline
		% \multirow{1}{*}{\rotatebox[origin=c]{0}{\makecell{ ~R~ }}}
		$ J_\mathrm{R}^\mathrm{S}$ 
		& $B^2\tauS^4 + B\left(4+3B\right)\tauS^3 + 2B(6+B)\tauS^2
		+B(8+B)\tauS - 4(1-B) $ \\
		\hline\hline
		\end{tabular}
		\caption{Expressions of the time-dependent coefficients appearing in~Tab.~\ref{tab:time-dependent-expressions}.
		}
		\label{tab:coeff}
\end{center}
\end{table*}

\clearpage

% \bibliography{biblio,bibliodfg} % No space after comma!!

\end{document}